\documentclass[preprint, double-spaced]{revtex4}
\usepackage[colorlinks=true,linktocpage=true,linkcolor=blue,citecolor=blue]{hyperref}
\usepackage{graphicx}
\usepackage{amsmath,bbm}
\usepackage{amssymb,bm}
\usepackage{array}
\usepackage{float}

\begin{document}
\title{Transverse Momentum Distribution and Elliptic Flow of Charged Hadrons in $U$+$U$ collisions at $\sqrt{s_{NN}}=193$ GeV using HYDJET++}
\author{Arpit Singh$^1$}
\author{P. K. Srivastava$^{2,}$\footnote{prasu111@gmail.com,~prashant.srivastava@iitrpr.ac.in}}
\author{O. S. K. Chaturvedi$^1$}
\author{S. Ahmad$^{3}$}
\author{B. K. Singh$^{1,}$\footnote{bksingh@bhu.ac.in}}
\affiliation{$^1$Department of Physics, Institute of Science, Banaras Hindu University, Varanasi-221005, INDIA}
\affiliation{$^2$Department of Physics, Indian Institute of Technology Ropar, Rupnagar-140001, INDIA}
\affiliation{$^3$Department of Physics, Aligarh Muslim University, Aligarh-202002, INDIA}

\begin{abstract}
Recent experimental observations of the charged hadron properties in $U+U$ collisions at $193$ GeV contradict many of the theoretical models of particle production including two-component Monte Carlo Glauber model. The experimental results show a small correlation between the charged hadron properties and the initial geometrical configurations (e.g. body-body, tip-tip etc.) of $U+U$ collisions. In this article, we have modified the Monte Carlo HYDJET++ model to study the charged hadron production in $U+U$ collisions at $193$ GeV center-of-mass energy in tip-tip and body-body initial configurations. We have modified the hard as well as soft production processes to make this model suitable for $U+U$ collisions. We have calculated the pseudorapidity distribution, transverse momentum distribution and elliptic flow distribution of charged hadrons with different control parameters in various geometrical configurations possible for $U+U$ collision. We find that HYDJET++ model supports a small correlation between the various properties of charged hadrons and the initial geometrical configurations of $U+U$ collision. Further, the results obtained in modified HYDJET++ model regarding $dn_{ch}/d\eta$ and elliptic flow ($v_{2}$) suitably matches with the experimental data of $U+U$ collisions in minimum bias configuration. 
\end{abstract}

\maketitle 
\section{Introduction}
\noindent
The basic motivation of heavy ion collision experiments is to understand the properties and behaviour of quantum chromodynamics (QCD) at very high temperature and chemical potentials via analysing the data on multi-particle production and by matching experimental measurements to the simulation models for the entire evolution of the fireball. There are existing computational models which use the theoretical or phenomenological foundation of strong interactions to mimic the space-time evolution of collision experiments. One can broadly classify these models in two types: dynamical models ~\cite{Schenke:2012,Eskola:2000,Eskola:2002,Kharzeev:2004,Kharzeev:2001,Schee:2013,Berges:2014,Kurkela:2014} and semi dynamical models ~\cite{Miller:2007,Lin:2005,Bleicher:1999}. Dynamical models are those which consider the pre-equilibrium evolution as well as post equilibrium hydrodynamic evolution like IP-Glasma model etc~\cite{Schenke:2012,Eskola:2000,Eskola:2002,Kharzeev:2004,Kharzeev:2001,Schee:2013,Berges:2014,Kurkela:2014}. However, most of the models are semi dynamical models which use a static initial condition at proper thermalization time and then evolve the system using viscous or ideal hydrodynamics like AMPT, MC-Glauber etc~\cite{Miller:2007,Lin:2005,Bleicher:1999}. The particle production mechanism of both types of model are quite different. In dynamical models, the parton saturation is a viable mechanism for particle production e.g., IP-Glasma model is based on the ab-initio color glass condensate framework which combines the impact parameter dependent saturation model for parton distributions with an event-by-event classical Yang-Mills description of early-tile glasma fields~\cite{Schenke:2012}. Similarly EKRT model is based on the assumption of final state gluon saturation and thus the initial energy density and produced number of partons scales with atomic number and beam energy~\cite{Eskola:2000,Eskola:2002}. In KLN model, the inclusive production of partons is driven by the parton saturation in strong gluon fields~\cite{Kharzeev:2004,Kharzeev:2001}. In saturation regime, the multiplicity of produced partons should be proportional to atomic number~\cite{Kharzeev:2004,Kharzeev:2001}. On the other hand the particle production mechanism in semi-classical models are implemented via some phenomenological parameterization or using Monte Carlo event generator e.g., in MC-Glauber model, the particle production is based on static initial conditions and two-component parameterization in which first term is proportional to mean number of participants and second term is proportional to mean number of collisions~\cite{Miller:2007}. In AMPT model initial conditions are obtained from HIJING event generator then ZPC for parton scatterings. After that Lund string model for hadronization and ART model to treat the hadronic scatterings~\cite{Lin:2005}. UrQMD model describes the particle production at low and intermediate energies in terms of scatterings amongst hadrons and their resonances. At higher energies, the excitation of colour strings and their subsequent fragmentation is the particle production mechanism in this model~\cite{Bleicher:1999}.

Most of the simulation models are successful in providing the multiplicity of charged hadrons produced in various heavy ion collision experiments. Vast experimental data on multi-particle production and distributions with collision control parameters like centrality, rapidity and/or transverse momentum etc., put a stringent constraint on these models so that we can understand the production mechanism more deeply and make our models more realistic. To strengthen our understanding about quantum chromodynamics (QCD), these collider experiments collide various nuclei at different colliding energies. Recently RHIC experiment has collided uranium ($U$) nuclei at the center-of-mass energy $\sqrt{s_{NN}}= 193$ GeV ~\cite{Pandit:2014}. As we know that uranium is a deformed nuclei (prolate in shape) so various kind of initial configurations are possible in $U+U$ collision e.g., body-body, tip-tip, body-tip etc.  The various computational models previously predicted a large difference in multiplicity and elliptic flow between body-body and tip-tip configurations of $U+U$ collisions ~\cite{Bjorn:1403,Sergei:1006}. However, the experimental data of multi-particle production in $U+U$ collisions regarding multiplicity and elliptic flow ($v_{2}$) contradicts the earlier expectations of most of these computational and theoretical models and shows a small correlation between multiplicity (and/or $v_{2}$) and initial configurations of $U+U$ collision ~\cite{Pandit:2014}. This contradiction may have two possible reasons. Either the simulation models have something missing or experimentally we are not quite able to disentangle the events with different geometrical orientations. Thus we have to work on both the aspects since $U+U$ collision in its various orientations is quite useful to understand wide range of physics. Quark gluon plasma (QGP) phase which is characterized by the observables like elliptic flow, jet quenching, charmonia suppression and multiplicity can be better understood in the collision of deformed uranium nuclei due to its initial geometry and specific orientation ~\cite{Hiroshi:2009,Shou:2015,Sergei:1006,Gold:2015,Hirano:2011,Kuhlman:2005}.  Further $U+U$ collisions can provide a reliable tool to subtract the background elliptic flow effect from the signal so that one can detect the chiral magnetic effect (CME)~\cite{Sergei:1006}. In spherical nuclei, it is difficult to disentangle both these effect since the strength of both the signals generated from elliptic flow and CME is of similar strength in peripheral collisions. However, in $U+U$ central collisions, the different geometrical orientations can provide a way to subtract the background signal from CME signal due to a measurable difference in their strength. Thus central collisions of $U+U$ nuclei in tip-tip configuration can possibly be a good tool to characterize the signal of CME ~\cite{John:1311,Bjorn:1403}. 

Very recently different methods have been proposed to modify some of the models to incorporate the experimental $U+U$ observations in that particular simulation models~\cite{Chatterjee:2016,Rybcz:2013,Moreland:2015}.  The constituent quark model is also proposed to describe the experimental observation on $v_{2}$ in $U+U$ collisions~\cite{Eremin:2003,Adler:2014}. In this article we want to study the $U+U$ collision at $\sqrt{s_{NN}} = 193$ GeV in body-body and tip-tip configurations by modifying HYDJET++ model which uses PYTHIA type initial condition for hard part and Glauber type initial condition for soft part. Further most of the existing models either consist of high $p_{T}$ particle production from jet fragmentation or involve low $p_{T}$ hadron production using thermal statistical processes. However, HYDJET++ model~\cite{Lokhtin:2009}  consistently includes production of hard as well as soft $p_{T}$ hadrons, to calculate the charged hadron production in $U+U$ collisions at center-of-mass energy $\sqrt{s_{NN}}=193$ GeV. We study the pseudorapidity distribution, transverse momentum ($p_{T}$) distribution of charged hadrons. Moreover we calculate the elliptic flow of these produced particles in body-body and tip-tip configurations of $U+U$ collisions. Rest of the article is organised as follows: In section II, we have provided a brief detail of modified HYDJET++ model and described its various physical parts under different subsections. Further we have written down the equation to calculate the elliptic flow of charged hadrons. In Section III, we have provided the results and discussions under two subsections: (A) pseudorapidity distributions and, (B) transverse momentum distribution and elliptic flow. At last we have summarized our current work.

\section{Model Formalism}
The heavy ion event generator HYDJET++ simulates relativistic heavy ion collisions as a superposition of the soft, hydro-type state and the hard state resulting from multi-parton fragmentation. The soft and hard components are treated independently in HYDJET++. The details on physics model and simulation procedure of HYDJET++ can be found in the corresponding manual ~\cite{Lokhtin:2009,Bravina:2017}. The main features of HYDJET++ model are listed very briefly in this section.\\
\subsection{Hard multi-jet production}
The model for the hard multi-parton production of HYDJET++ event is based on PYQUEN partonic loss model ~\cite{Lokhtin:2006,Lokhtin:2708,Lokhtin:2082}. In brief the hard part of hadron production in HYDJET++ uses PYQUEN ~\cite{Lokhtin:2006} which includes generation of initial parton spectra according to PYTHIA and production vertices is measured at a given impact parameter. After that rescattering of partons is incorporated using an algorithm of the parton path in a dense medium along with their radiative and collisional energy loss. Finally hadronization takes place according to the Lund string model ~\cite{Anderson:1998} for hard partons and in-medium emitted gluons. An important cold nuclear matter effect which is shadowing of parton's distribution function is included using Glauber-Gribov theory ~\cite{Gribov:1969}.
As a simplification to the model, the collisional energy loss due to scattering~\cite{sve,pks_drag} with low momentum transfer is not considered because its contribution to the total collisional energy loss is very less in comparison with high momentum scattering. The medium where partonic rescattering occurs is treated as a boost invariant longitudinally expanding quark-gluon fluid, and partons are produced on a hypersurface of equal proper times $\tau$ ~\cite{Bjorken:1983}. Since we use Bjorken hydrodynamics thus the results in this model have limited applicability at larger rapidities where one should use Landau hydrodynamics for the proper description of medium expansion. 
\begin{figure}
\includegraphics[scale=0.70]{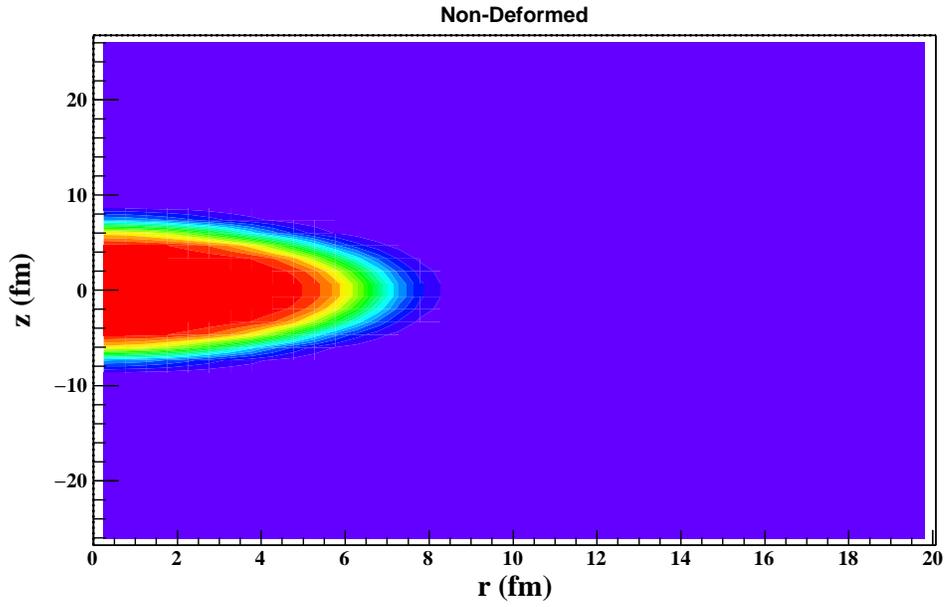}
\caption{(Color online) Nuclear density contour in $r-z$ plane of cylindrical coordinate system for non-deformed nucleus. We follow the VIBGYOR colour coding in this figure. Red means the highest value of nuclear density and violet represents the lowest nuclear density.}
\end{figure}

\begin{figure}
\includegraphics[scale=0.70]{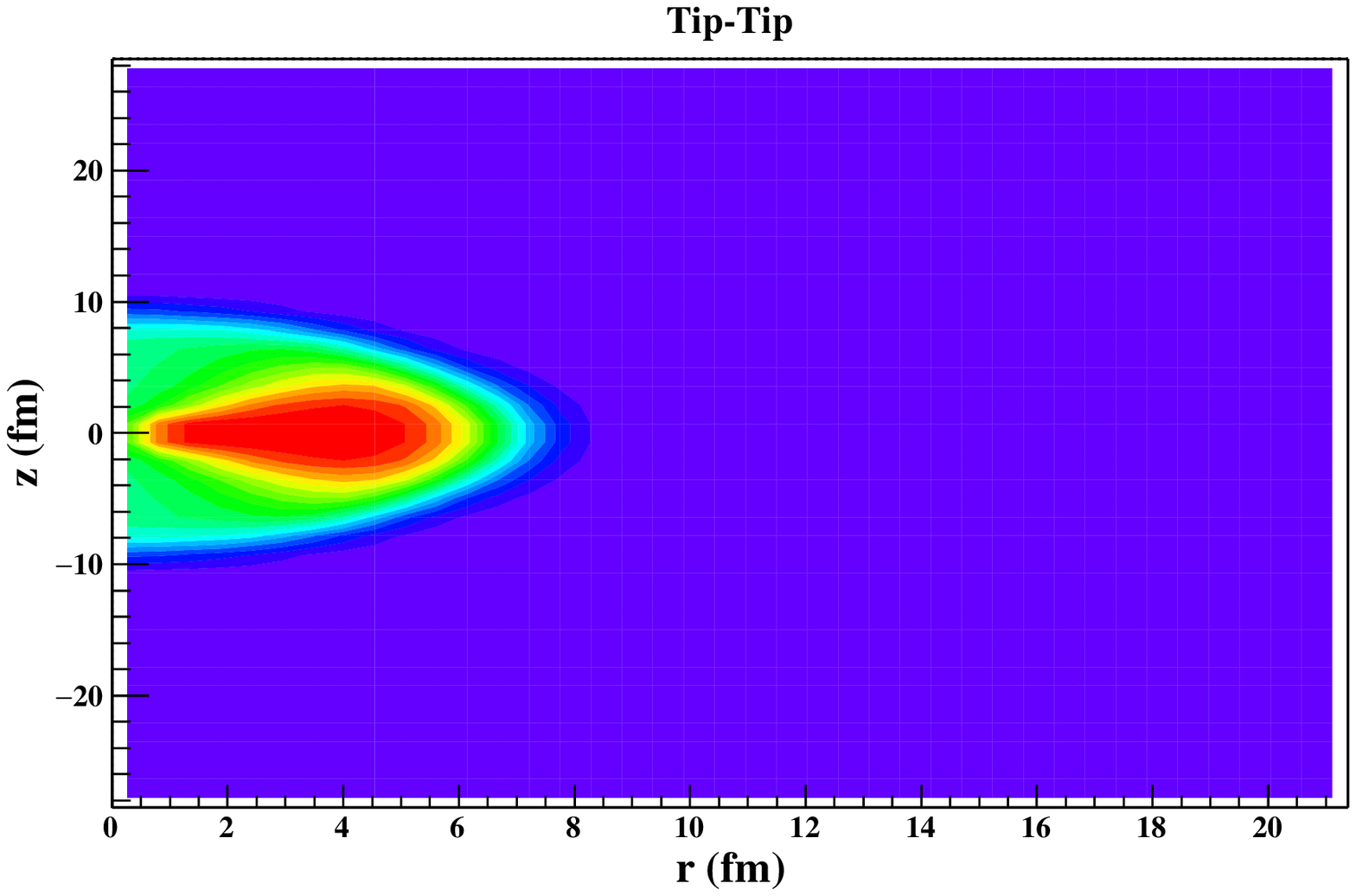}
\caption{(Color online) Nuclear density contour in $r-z$ plane of cylindrical coordinate system for uranium nucleus in tip configuration. Representation of colours is same as in the case of Fig. 1.}
\end{figure}
\begin{figure}
\includegraphics[scale=0.70]{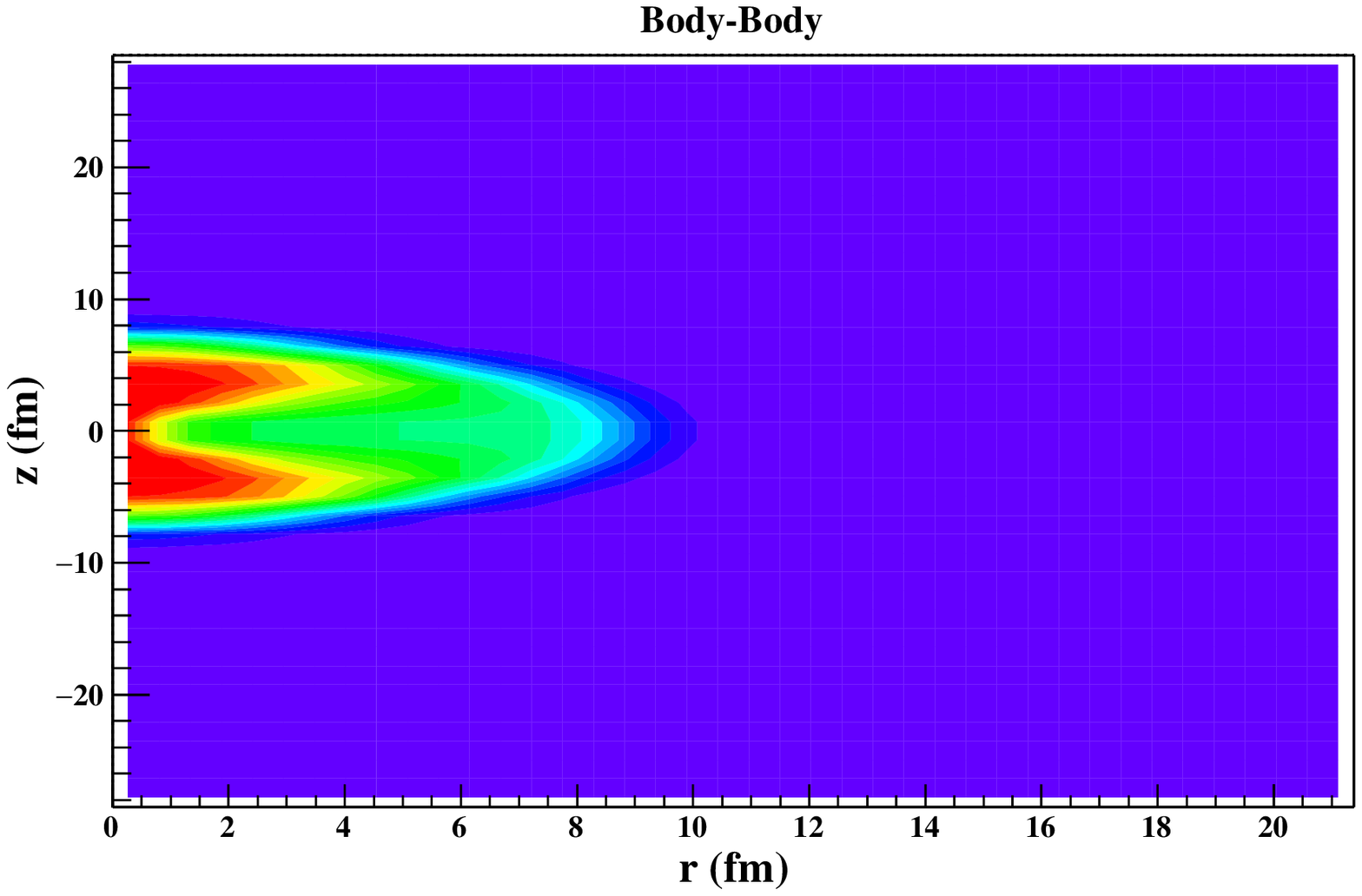}
\caption{(Color online) Nuclear density contour in $r-z$ plane of cylindrical coordinate system for uranium nucleus in body configuration. Representation of colours is same as in the case of Fig. 1.}
\end{figure}

The main modification which we have done in the present version of HYDJET++ is, to change the nuclear density profile function. However, this modification is not straightforward in HYDJET++ as done in AMPT by other authors~\cite{Rihan:2012} since HYDJET++ deals in cylindrical polar coordinates ($\rho,~z,~\psi$) instead of spherical polar coordinate system ($r,~\theta,~\phi$). To make HYDJET++ work for $U+U$ collisions, one has to transform the deformed Woods-Saxon nuclear density profile function from spherical polar to cylindrical polar coordinate system. In spherical polar coordinates the deformed Woods-Saxon for uranium nucleus is defined as follows~\cite{OSK:2016}:
\begin{equation}
  \rho(r,z,\theta) = \rho_{0}\frac{1}{1+\exp(\frac{r -R(1+\beta_{2}Y_{20}+\beta_{4}Y_{40})}{a})}   
\end{equation}
where $\rho_{0}$ is calculated using a simple equation i.e., $\rho_{0}=\rho_{0}^{const}+correction$ and $\rho_{0}^{const}=~Mass/Volume = A/(4\pi R_{A}^{3}/3)=3/(4\pi R_{l}^{3})$, where radius of uranium nucleus $R_{A}=R(1+\beta_{2}Y_{20}+\beta_{4}Y_{40})$, $R_{l}=R_{0}(1+\beta_{2}Y_{20}+\beta_{4}Y_{40})$, and $R=R_{0}A^{1/3}$ with $R_{0}=1.15$ fm. The correction term is calculated by using $\rho_{0}^{const}$ as $correction =~\rho_{0}^{const}\times (\pi~f/R_{A})^{2}$ with $f=0.54$ fm. Further $Y_{20} = \sqrt{\frac{5}{16\pi}}(3\cos^{2}(\theta)-1)$ , $Y_{40} = \frac{3}{16\sqrt{\pi}}(35\cos^{4}(\theta)-30\cos^{2}(\theta)+3)$ are the spherical harmonics with the deformation parameters $\beta_{2}$ and $\beta_{4}$.The different parameter values for uranium nuclei are taken from Refs. ~\cite{Loizides:2549,Shou:2015}. Here the body-body and tip-tip configuration is mainly controlled by $\theta$ and all other coordinates integrated over same range. However as shown in Ref.~\cite{Rihan:2012} one can change the range of $\phi$ to make other configurations as well but here we will stick to body-body and tip-tip configurations. In the conversion of nuclear density profile from spherical polar to cylindrical polar coordinate, we find a relation $\theta=tan^{-1}(r/z)$ and $\theta=tan^{-1}(z/r)$ for tip-tip and body-body configuration of $U+U$ collision, respectively. Here $r$ is basically $\rho$ of cylindrical polar coordinate system and not spherical polar coordinate $r$. We follow this representation so that readers do not get confused it with nuclear density function ($\rho$). The values and range of $\psi$ remains equal to $\phi$ during this coordinate conversion as far these two configurations are concerned. It is quite difficult to make conversion mapping between these two coordinate systems to incorporate random values of theta from its whole range i.e. $0$ to $\pi$. Thus we reserve this topic for our future research work. To show the validity of our modification in deformed Woods-Saxon function and make the readers visualize, the nuclear density profiles (in cylindrical coordinate system) for non-deformed gold nucleus along with tip and body configuration of uranium nucleus are shown in Fig. 1, 2 and 3, respectively. 
\begin{figure}
\includegraphics[scale=0.40]{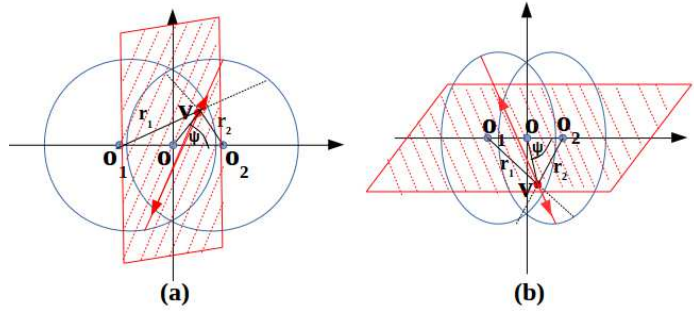}
\caption{(Color online) Jet production for (a) tip-tip configuration, and (b) body-body configuration of U+U collisions in the plane of impact parameter $b$. $O_{1}$ and $O_{2}$ are the nucleus centers, $OO_{1}=OO_{2}=b/2$. V is the jet production vertex and its coordinate will be ($r~cos\psi ,~r~sin\psi$) for tip-tip and body-body configuration both.}
\end{figure}
Now the two quantities, nuclear thickness function ($T_{A}$) and nuclear overlap function ($T_{AA}$) can be calculated using this modified and deformed Woods-Saxon nuclear density profile function in cylindrical coordinates $\rho (r,~z,~\psi)$ by following expressions~\cite{Lokhtin:2000}(Please see Fig. 4 (a) and (b)):
\begin{eqnarray}
  T_{AA}(b) &=& \int_{0}^{\infty}r~drd\psi~T_{A}(r_{1})T_{A}(r_{2}) \ 
\end{eqnarray}
\begin{equation}
  T_{A}(r) = A\int\rho_{A}(r,z,\psi)dz \  ,\   r_{1,2} = \sqrt{r^{2}+\frac{b^{2}}{4}\pm rb\cos(\psi)}
\end{equation}
where $r_{1,2}(b,r,\psi)$ are the distances between the centers of colliding nuclei and the jet production vertex $V(r\cos\psi,r\sin\psi)$, $r$ is the distance from the nuclear collision axis to $V$, $R_{eff}(b,\psi)$ is the transverse distance from the nuclear collision axis to the effective boundary of nuclear overlapping area in the given azimuthal direction $\psi$.\\
To further validate the predictive power of our model for tip-tip and body-body configurations of $U+U$ collisions, we have calculated the number of participants ($N_{part}$) and number of binary collisions ($N_{coll}$) in both the configurations and plotted them on Fig. 5 with respect to ratio of impact parameter ($b$) with maximum possible radius of uranium nucleus ($R_{A}$). This $b/R_{A}$ actually represents the centrality of the event. We have also shown the maximum possible value of $N_{part}(0)$ for both the configurations which must not depend on centrality. 

\begin{figure}
\includegraphics[scale=0.70]{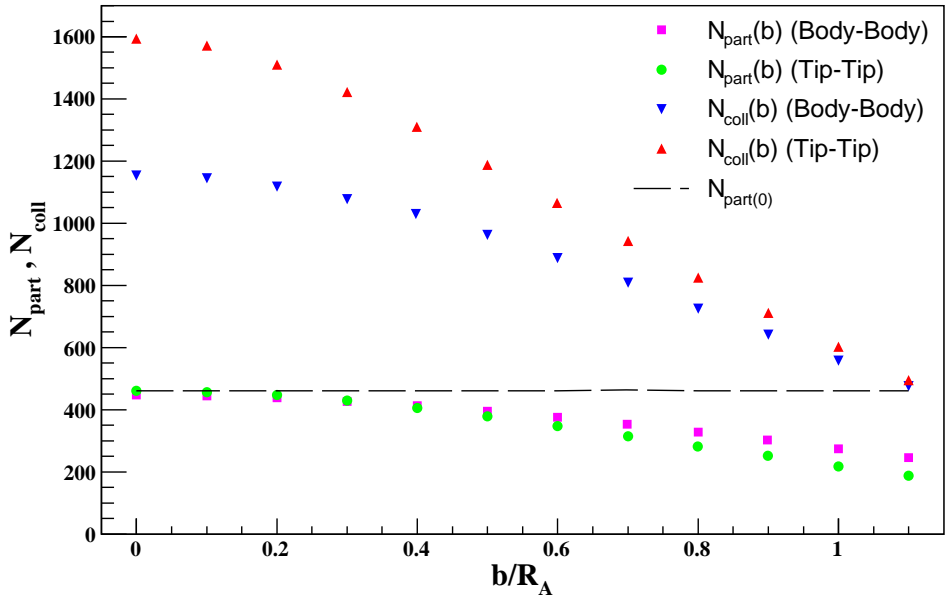}
\caption{(Color online) Variation of $N_{part}$ and $N_{coll}$ with impact parameter for tip-tip and body-body configuration of $U+U$ collisions.}
\end{figure}

\subsection{Soft 'thermal' hadron production}
The soft part of HYDJET++ is the thermal hadronic state generated on the chemical and thermal freeze-out hypersurface obtained from the parameterization of relativistic hydrodynamics with a given freeze-out condition~\cite{Amelin:2006,Amelin:2008}. The first and foremost modification which we have done in soft part is to change the nuclear density profile function for deformed uranium nucleus as discussed in above subsection. After that we have to modify the freeze-out hypersurface to properly include the effect of nuclear deformation via change in number of participants.

There are various ways to generate the initial conditions for chemical and thermal freeze-out hypersurface~\cite{cleymans,skt,sandeep}. However we first want to start here with the hydrodynamic evolution of this freeze-out hypersurface i.e., the hydrodynamic evolution laws for QCD medium. In HYDJET++, the QCD medium is assumed to evolve according to the Bjorken boost-invariant hydrodynamics. Therefore the cooling laws for energy density and temperature are as follows~\cite{Lokhtin:2000}:
\begin{equation}
\epsilon (\tau) \tau^{4/3}=\epsilon_{0}\tau_{0}^{4/3}, and
\end{equation}  
\begin{equation}
T(\tau) \tau^{1/3}=T_{0}\tau_{0}^{1/3},
\end{equation}
respectively. In above equations, $\epsilon_{0}$, and $T_{0}$ are the initial energy density, and temperature at initial proper time $\tau_{0}$ at which the local thermal equilibrium has been established. The initial energy density at $\tau_{0}$ and at impact parameter $b=0$ is calculated by estimating the energy density inside the co-moving volume of longitudinal size i.e., $\Delta z$ for tip-tip and $\Delta r$ for body-body configuration. The expression of total initial transverse energy deposition in the mid-rapidity region is as follows\cite{Lokhtin:2000}:
\begin{equation}
\epsilon_{0}(b=0,\tau_{0}) = T_{AA}(0).\sigma_{NN}^{jet}(\sqrt{s},p_{0}).\langle p_{T}\rangle,
\end{equation}
where $T_{AA}$ can be calculated by using Eq. (2) and (4) for tip-tip and body-body, respectively. $\sigma_{NN}^{jet}(\sqrt{s},p_{0}).\langle p_{T}\rangle$ is the first $p_{T}$ moment of the inclusive differential minijet cross-section which is determined by the dynamics of the nucleon-nucleon interactions at the corresponding c.m.s. energy. The initial energy density at a given impact parameter can be calculated from the following expression\cite{Lokhtin:2000}:
\begin{equation}
\epsilon_{0}(b,\tau_{0})=\epsilon_{0}(b=0,\tau_{0}).\frac{T_{AA}(b)}{T_{AA}(0)}.\frac{S_{AA}(b)}{S_{AA}(0)},
\end{equation}
where $S_{AA}(b)$ is effective transverse area of the nuclear overlapping zone at impact parameter $b$~\cite{Lokhtin:2000} and is calculated as:\begin{equation}
  S_{AA}(b)=\int_{0}^{2\pi}d\psi\int_{0}^{r_{max}}~rdr
\end{equation}
Now to calculate the initial temperature in our calculations we have used a parameterization based on ideal thermal gas approximation~\cite{skt,andronic} where $T_{0}(b=0,\tau_{0})$ and baryon chemical potential $\mu_{0}(b=0,\tau_{0})$ can be calculated from the collision energy using the following relations:
\begin{equation}
\mu_{0}(b=0,\tau_{0})= \frac{a}{1+b\sqrt{s_{NN}}},
\end{equation}
\begin{equation}
T_{0}(b=0,\tau_{0}) = c-d\mu_{B}^{2} - e\mu_{B}^{4}.
\end{equation}
Here the parameters $a,~b,~c,~d,$ and $e$ have been determined from the best fit of the particle ratios at various collision energies: $a =1.290\pm 0.113$ GeV, $b=0.28\pm 0.046$ GeV$^{-1}$, $c= 0.170\pm 0.1$ GeV, $d=0.169\pm 0.02$ GeV$^{-1}$, and $e=0.015\pm 0.01$ GeV$^{-3}$. 
The temperature for other i.e., semi-central, semi-peripheral and peripheral, events is calculated by using the following relation so that one can convert the fixed freeze-out hypersurface into a centrality(or $N_{part}$) dependent hypersurface which is much needed modification in soft particle production in HYDJET++:
\begin{equation}
T_{0}(b,\tau_{0}) = T_{0}(b=0,\tau_{0})\times\left(\frac{N_{part}(b)}{N_{part}(0)}\right)^{1/3}.
\end{equation}
We have treated the $\mu_{B}$ as centrality independent since the value of baryon chemical potential is small at highest RHIC energies and thus the effect of change due to centrality dependence should not affect the multiplicity by more than 5$\%$~\cite{Lokesh}. Further hadron multiplicities are calculated using the effective thermal volume approximation and Poisson multiplicity distribution around its mean value, which is supposed to be proportional to the number of participating nucleons at a given impact parameter of A-A collision. We have shown the change in effective thermal volume between body-body and tip-tip configuration with respect to $b/R_{A}$ in Fig. 6. We have also plotted the variation of chemical freeze-out temperature with respect to $b/R_{A}$ on the same plot (Fig. 6) for body-body and tip-tip configuration of $U+U$ collision at $\sqrt{s_{NN}}=193$ GeV. Feed-down corrections from two- and three-body decays of the resonances with branching ratios are taken according to SHARE particle decay table ~\cite{Torrieri:2005} when calculating the final multiplicity of the particles. 
\begin{figure}
\includegraphics[scale=0.70]{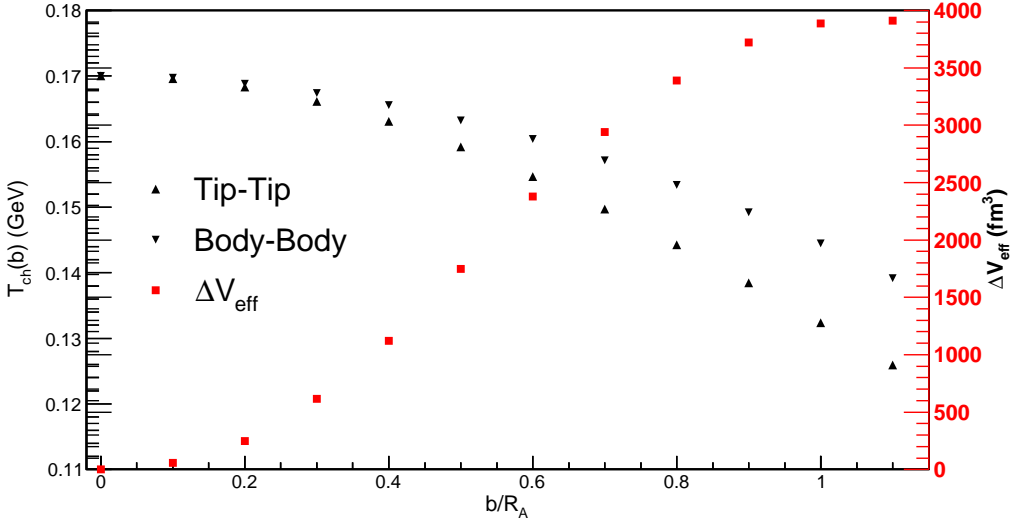}
\caption{(Color online) Variation of $T_{ch}(b)$ and $\Delta V_{eff}$ with impact parameter for tip-tip and body-body configuration of $U+U$ collisions.}
\end{figure} 
\subsection{Elliptic flow}
Non-central collisions generate an initial spatial asymmetry of almond shape in the plane transverse to the reaction plane. The re-interactions among the reaction products in the initial state converts this spatial anisotropy into particle momentum anisotropy. In other words the spatial anisotropy in the collision zone results in anisotropic pressure gradients that generate stronger (weaker) collective flow in the direction of the major (minor) axis of the almond-shaped reaction zone. This phenomenon is called elliptic flow and is measured by $v_{2}$. The elliptic flow coefficient $v_{2}$ is determined as the second-order Fourier coefficient in the hadron distribution over the azimuthal angle $\psi$ relative to the reaction plane $\psi_{R}$~\cite{Bravina:2017}, so that
\begin{equation}
  v_{2} = <\cos2(\psi-\psi_{R})>
\end{equation}
Here,
\begin{equation}
   \psi=\tan^{-1}(p_{y}/p_{x})
\end{equation}
In HYDJET++ framework, the reaction plane of order two is zero for all the events. The above Eq. (12) can be rewritten in a simpler form as follows~\cite{Eyyubova:2009}
\begin{equation}
  v_{2} = \langle\frac{p_{x}^{2}-p_{y}^{2}}{p_{x}^{2}+p_{y}^{2}}\rangle=\langle\frac{p_{x}^{2}-p_{y}^{2}}{p_{T}^{2}}\rangle.
\end{equation}
As we know that most of the elliptic flow arises due to the contribution of soft hadrons having lower transverse momentum and the role of hadrons having large transverse momentum is rather subdued. In HYDJET++ model, soft particle emission takes from a freeze-out hypersurface at the time of freezeout. Consequently, the elliptic flow arises in HYDJET++ model is not directly related to the initial spatial anisotropy ($\epsilon_{0}$) of the participating nucleons as it is in other models like AMPT etc.  In HYDJET++, we create a fireball which have geometrical irregularities in different directions of phase space at the time of freezeout and we assume that these irregularities are somewhat related with the initial spatial distribution of the participating nucleons in the collision region but in an involved manner. The shape of the fireball in the transverse region $x-y$ at the freezeout can be approximated by an ellipse in non-central collision. Radii $R_{x}$ and $R_{y}$ of the ellipse at a given impact parameter $b$ can be parameterized ~\cite{Retiere:2004,Huovinen:2001,Wiedemann:1998,Broniowski:2003} in terms of spatial anisotropy at freezeout $\epsilon_{2}(b) = (R_{y}^{2}-R_{x}^{2})/(R_{x}^{2}+R_{y}^{2})$ and the scale factor $R_{f}(b) = [(R_{x}^{2}+R_{y}^{2})/2]^{1/2}$ as:
\begin{equation}
  R_{x}(b) = R_{f}(b)\sqrt{1-\epsilon_{2}(b)} \  , \   R_{y}(b) = R_{f}\sqrt{1+\epsilon_{2}(b)}.
\end{equation}
The transverse radius $R_{ell}(b,\phi)$ of the fireball in the given azimuthal direction $\phi$ is related to spatial anisotropy at the time of freezeout as:
\begin{equation}
  R_{ell}(b,\phi) = R_{f}(b)\left(\frac{1-\epsilon_{2}^{2}(b)}{1+\epsilon_{2}(b)\cos2\phi}\right)^{1/2},
\end{equation}
where
\begin{equation}
  R_{f}(b) = R_{0}\sqrt{1-\epsilon_{2}(b)}.
\end{equation}
$R_{0}$ denotes the freeze-out transverse radius in central collision. 

\section{Results and Discussions}
\begin{figure}
\includegraphics[scale=0.70]{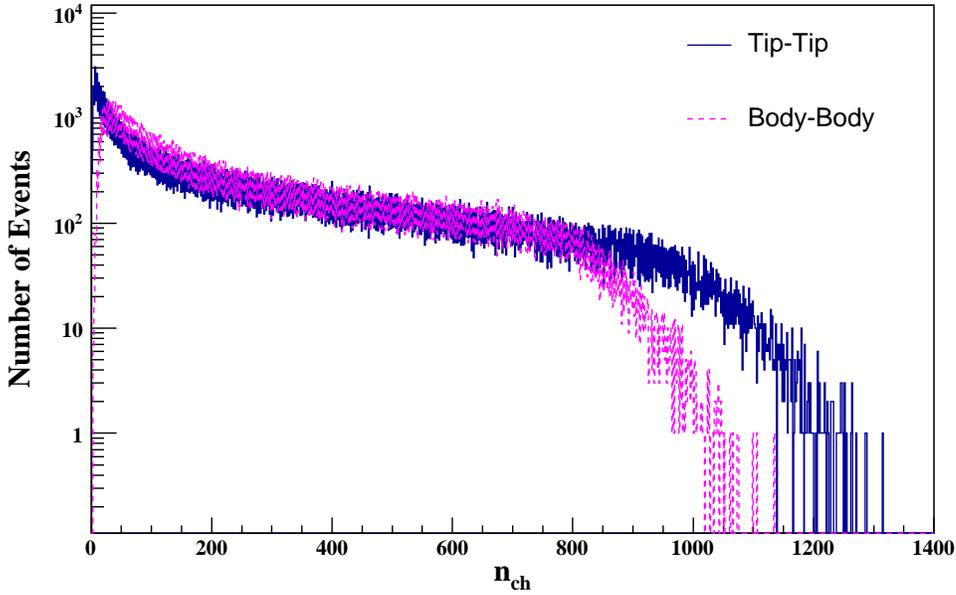}
\caption{(Color online) Variation of total charged particle multiplicity $(n_{ch})$ for tip-tip and body-body configuration of $U+U$ collisions.}
\end{figure}
\begin{figure}
\includegraphics[scale=0.70]{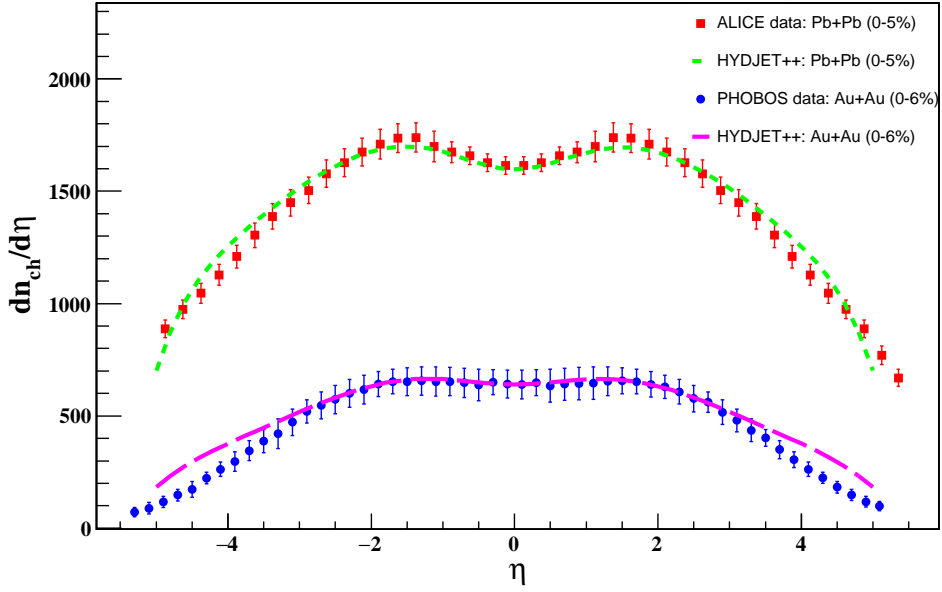}
\caption{(Color online) Variation of $dn_{ch}/d\eta$ with respect to $\eta$ is shown for $Au+Au$ and $Pb+Pb$ collisions in most central events. We have also plotted the corresponding experimental data~\cite{Adare:2016,Abbas:2013} for comparison.}
\end{figure}
\subsection{Pseudorapidity distributions}
\begin{figure}
\includegraphics[scale=0.50]{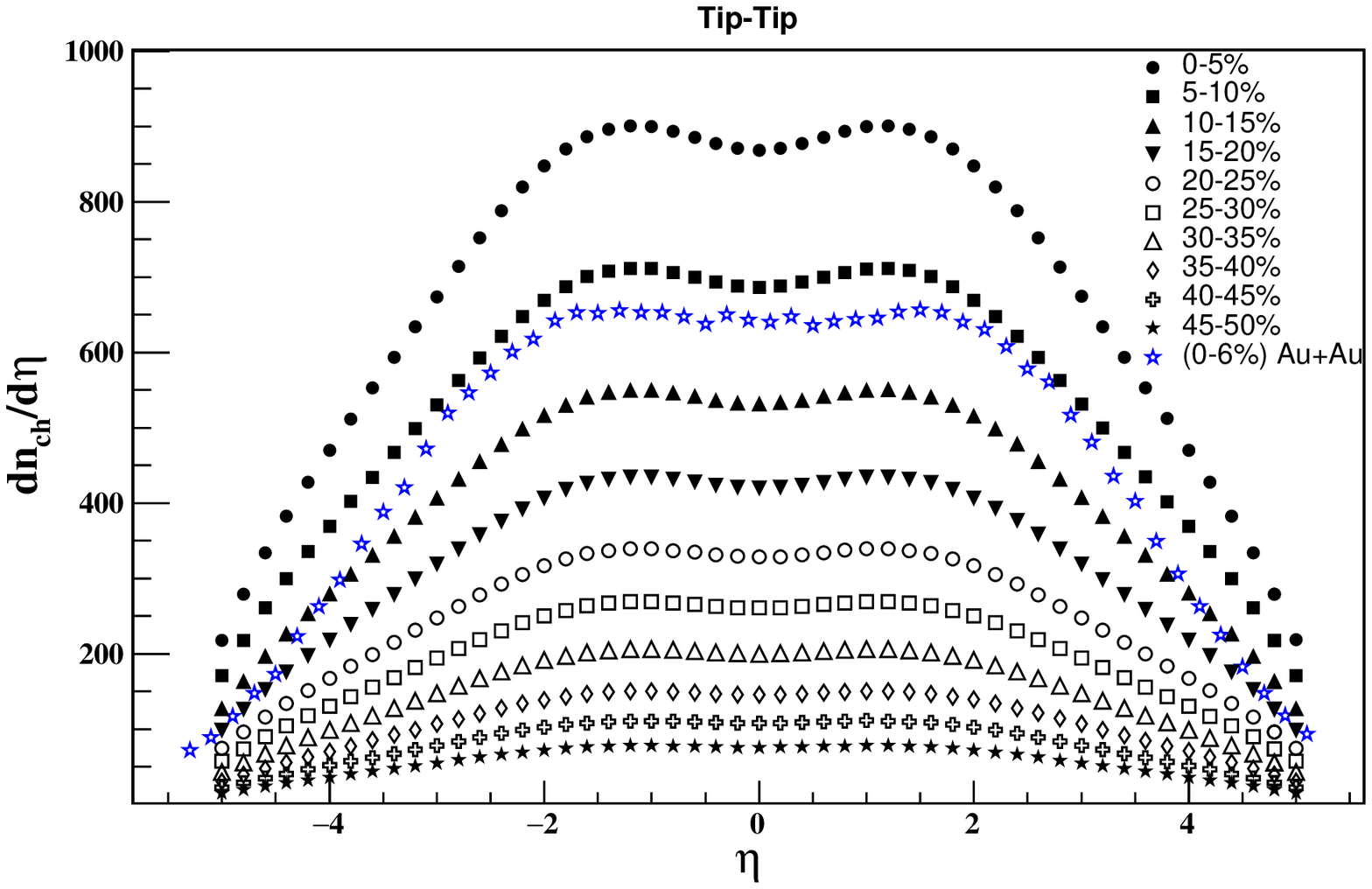}
\caption{(Color online) Variation of $dn_{ch}/d\eta$ with respect to $\eta$ is shown for tip-tip configuration of $U+U$ collisions in different centrality class.}
\end{figure}

\begin{figure}
\includegraphics[scale=0.50]{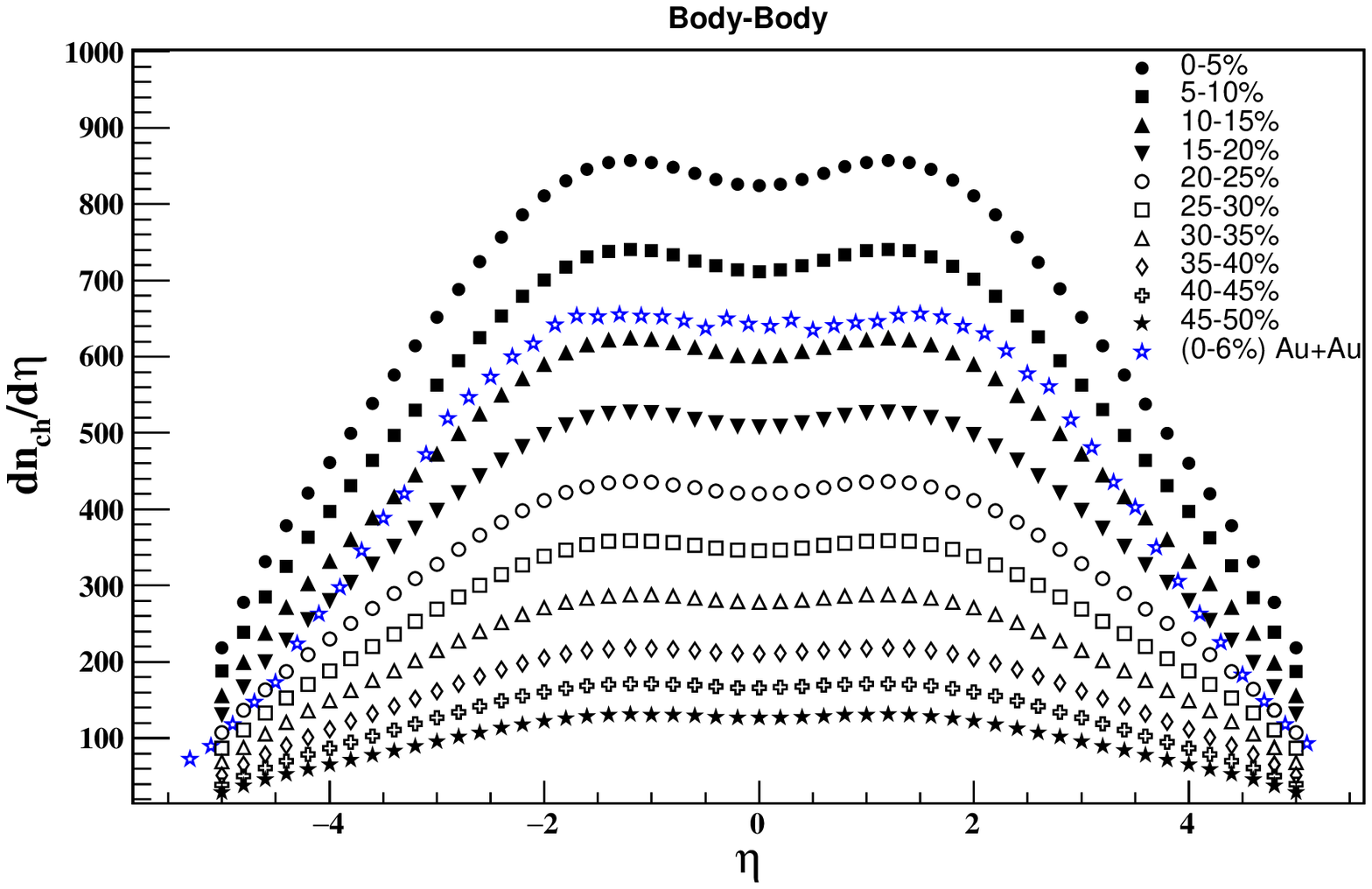}
\caption{(Color online) Variation of $dn_{ch}/d\eta$ with respect to $\eta$ is shown for body-body configuration of $U+U$ collisions in different centrality class.}
\end{figure}

\begin{figure}
\includegraphics[scale=0.50]{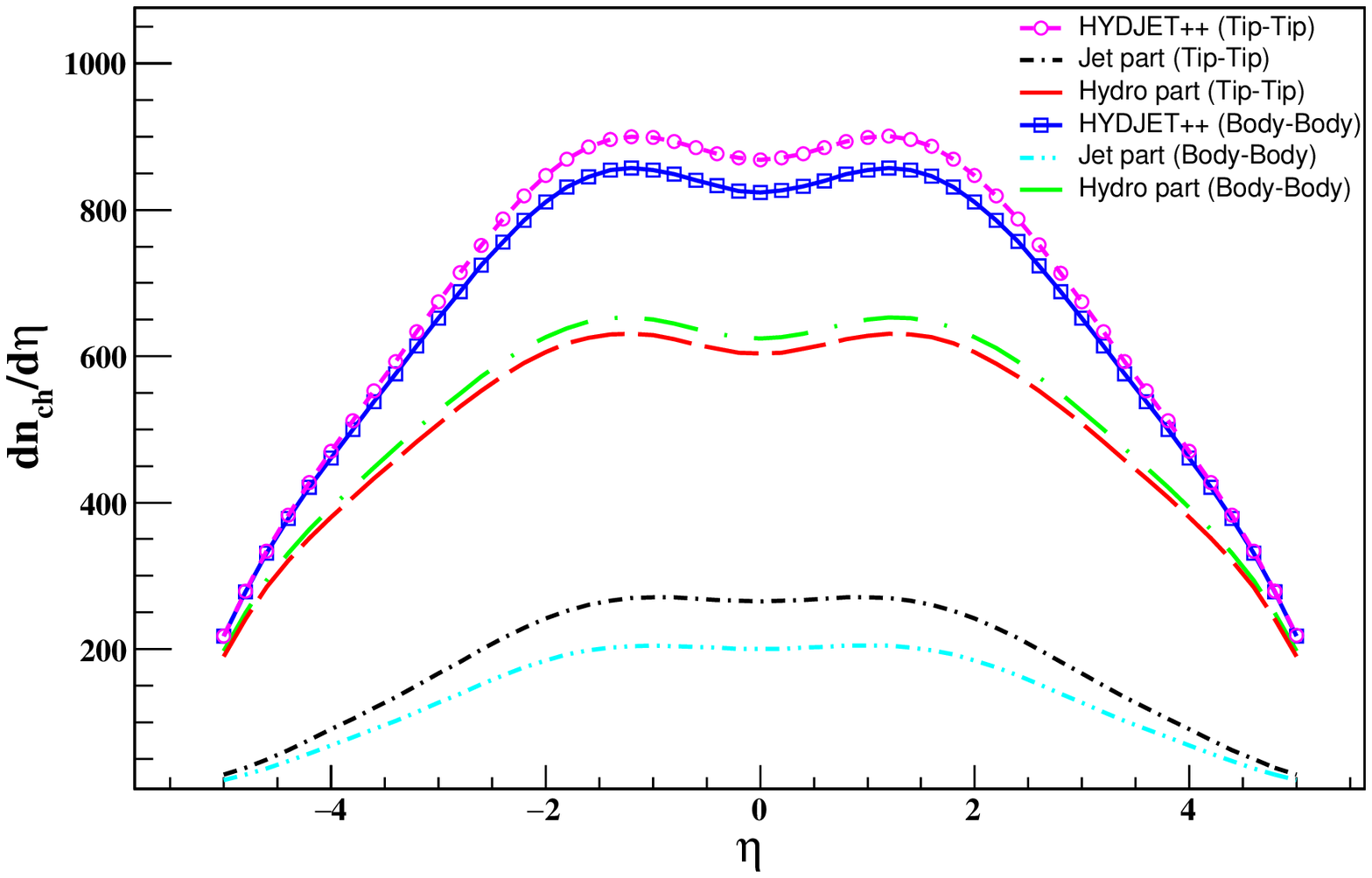}
\caption{(Color online) Comparison of pseudorapidity distribution in tip-tip and body-body configuration for most central collisions.}
\end{figure}

We have generated one million events for each centrality class for each of the configuration (tip-tip and body-body) separately using HYDJET++. Probability distribution curves for body-body and tip-tip events are shown in Fig. 7. We start our analysis with pseudorapidity distribution of charged hadrons. Pseudorapidity distribution of charged hadrons is a useful observable which can help us to understand various properties of the fireball formed and the particle production process.

In Fig. 8, we have plotted the pseudorapidity distributions of charged hadrons produced in $Au+Au$ and $Pb+Pb$ collisions at $200$ GeV and $2.76$ TeV for most central events and have compared HYDJET++ results with the published experimental data. This exercise authenticated the HYDJET++ model. Now we move towards the main aim of our study which is charged hadron production in $U+U$ collisions. In Fig. 9, we have shown the variation of $dn_{ch}/d\eta$ with respect to $\eta$ in tip-tip collisions of uranium nuclei at $\sqrt{s_{NN}}=193$ GeV. We have obtained these variations in various centrality intervals from most central ($0-5\%$) to most peripheral ($45-50\%$). The peak of these distributions has been occurred at $|\eta|=1.5$ and a little dip at $\eta=0$. The peak value of the number of charged hadrons is around $900$ in most central events and around $60$ in most peripheral events. Thus the increment is almost $15$ times in the number of produced charged hadrons ($dn_{ch}/d\eta$) at midrapidity going from peripheral to central tip-tip collisions.  We have also shown the experimental data points for particle multiplicity in $Au+Au$ collisions at $\sqrt{s_{NN}}=200$ GeV for most central events ~\cite{Back:2003}. One can see that $dn_{ch}/d\eta$ at midrapidity in most central $U+U$ collisions is larger than the most central $Au+Au$ collisions. Moreover it can be observed from the plot that the particle multiplicity in $5-10\%$ tip-tip configuration of $U+U$ collision is also larger than most central $Au+Au$ collision. The shape of distribution at larger rapidities is somewhat different in $U+U$ collision than $Au+Au$ collisions. However as we already mentioned in the model formulation section that HYDJET++ uses Bjorken boost invariant hydrodynamics which is not very much applicable at larger rapidities. Thus the observations at large rapidities may change if a proper hydrodynamical treatment is incorporated in HYDJET++ at large rapidities. In Fig. 10, we have presented the variation of pseudorapidty distribution with $\eta$ in body-body collisions between uranium nuclei. We have again presented the experimental multiplicity in most central $Au+Au$ collisions at $\sqrt{s_{NN}}=200$ GeV on this plot. Here again we found that the $dn_{ch}/d\eta$ at midrapidity in most central $U+U$ collisions is greater than $dn_{ch}/d\eta$ of most central $Au+Au$ collisions. 
\begin{figure}
\includegraphics[scale=0.50]{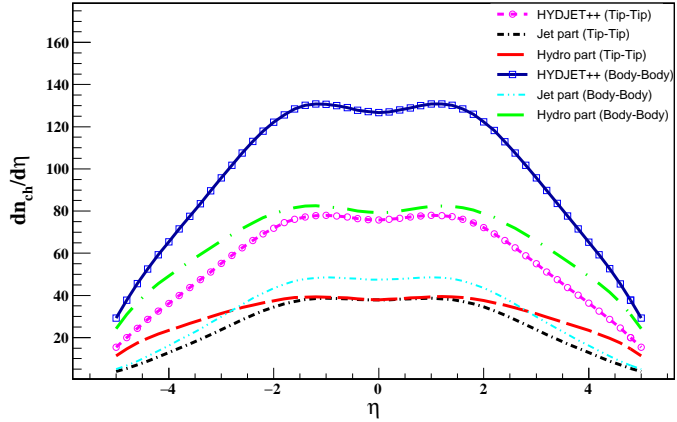}
\caption{(Color online) Comparison of pseudorapidity distribution in tip-tip and body-body configuration for most peripheral collisions.}
\end{figure}

\begin{figure}
\includegraphics[scale=0.50]{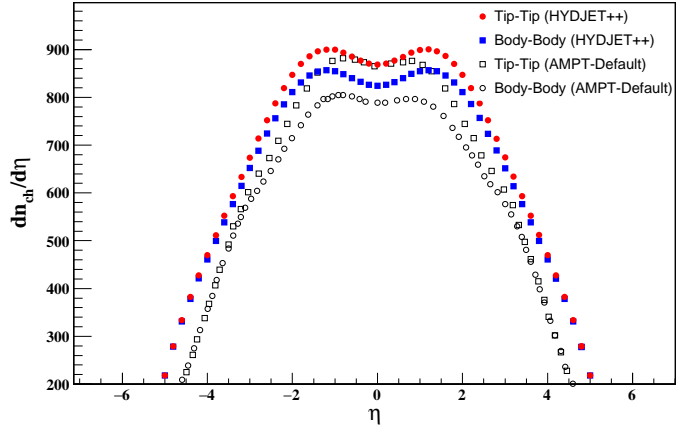}
\caption{(Color Online) Comparision of pseudorapidity distribution of HYDJET++ and AMPT model results in tip-tip and body-body configuration for most central collisions. AMPT results are taken from Ref.~\cite{Rihan:2012}.}
\end{figure}
\begin{figure}
\includegraphics[scale=0.50]{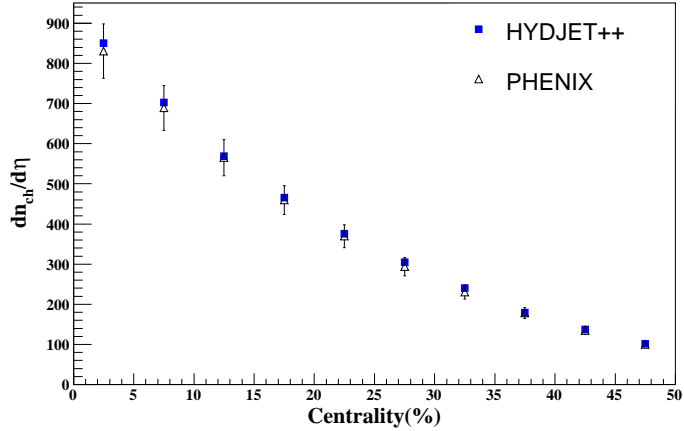}
\caption{(Color Online) Variation of $dn_{ch}/d\eta$ at midrapidity with centrality in minimum-bias configuration of $U+U$ collisions. Experimental data from PHENIX experiment ~\cite{Adare:2016} is also shown for comparison.}
\end{figure}

In Fig. 11, $dn_{ch}/d\eta$ with respect to $\eta$ is shown for most central tip-tip collision. Further we have presented the jet (hard) part and hydro (soft) part separately to show their relative contribution in the total multiplicity. From Fig. 11, one can see that the hard part has relatively low contribution than the soft part and hydro part is almost $3$ times larger than the jet part. One can also see that the jet part is almost flat in central rapidity region and the dip at $\eta = 0$ is mainly due to soft part of particle production. Further we have compared these most central tip-tip results with the most central body-body results. One can see that the combined multiplicity (soft plus hard) is larger in most central tip-tip configuration than the most central body-body configuration. Jet part has also the same behaviour. However, soft part shows an opposite behaviour. Here the body-body soft multiplicity is larger than tip-tip results. Similarly Fig. 12 presents the variation of $dn_{ch}/d\eta$ with respect to $\eta$ for most peripheral tip-tip configuration along with separate soft and jet part. Further we have compared these results with most peripheral body-body configuration. Here we found that the combined multiplicity is larger in body-body configuration than corresponding tip-tip result. Furthermore both jet as well as hydro part is larger in comparison to tip-tip configuration. Even the hydro part in body-body configuration is larger than the overall multiplicity in tip-tip configuration in most peripheral events.

In Fig. 13, we have compared our HYDJET++ results with the corresponding results obtained in AMPT model~\cite{Rihan:2012} for tip-tip and body-body configuration in $U+U$ collisions regarding $dn_{ch}/d\eta$. Authors of Ref.~\cite{Rihan:2012} have used $\sqrt{s_{NN}}=200$ GeV in their calculation however we have used $\sqrt{s_{NN}}=193$ GeV in our calculations. We found that the maxima in tip-tip configuration is similar in both models. However, the maxima in HYDJET++ model is larger than the maxima in AMPT model if body-body configuration is concerned. Thus the difference in multiplicity between body-body and tip-tip is smaller in HYDJET++ as compared to AMPT. Another difference between HYDJET++ and AMPT is the sharp decrease in $dn_{ch}/d\eta$ by increasing $\eta$ in AMPT as compared to HYDJET++ results. In Fig. 14, we have calculated the $dn_{ch}/d\eta$ at midrapidity in minimum bias configuration using HYDJET++. We have used a pseudorapidity cut as $|\eta|<0.5$. Further we have compared HYDJET++ results with the experimental results obtained by PHENIX collaboration~\cite{Adare:2016}. We found that the minimum bias data is successfully reproduced by HYDJET++ in the case of $dn_{ch}/d\eta$ at midrapidity.


\subsection{Transverse momentum distribution and elliptic flow}
\begin{figure}
\includegraphics[scale=0.70]{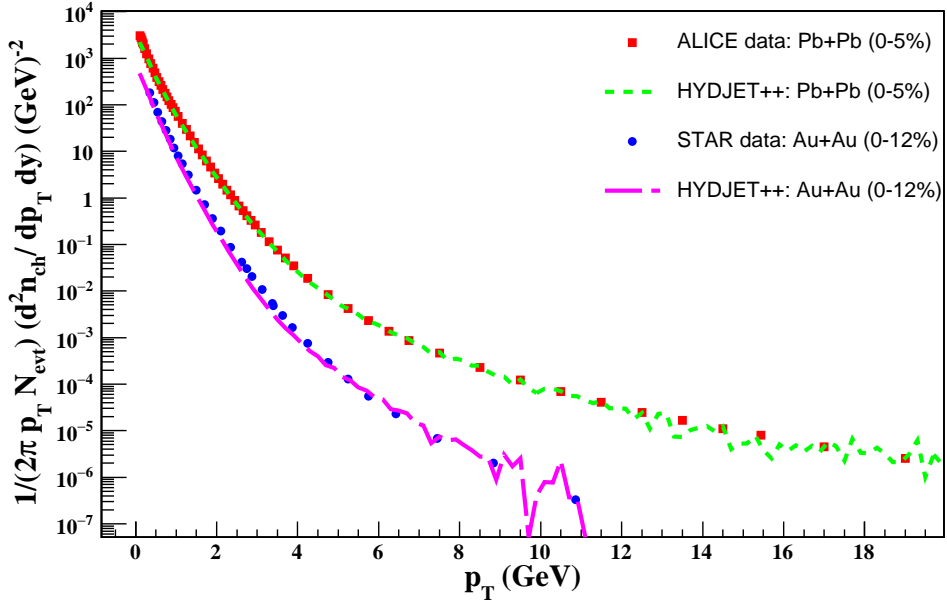}
\caption{(Color online) Variation of normalized transverse momentum distribution of positively charged pions in $Au+Au$ and charged pions $Pb+Pb$ collisions for most central class. We have plotted the experimental data from Refs.~\cite{Abelev:2006,Abelev:2014}.}
\end{figure}
\begin{figure}
\includegraphics[scale=0.50]{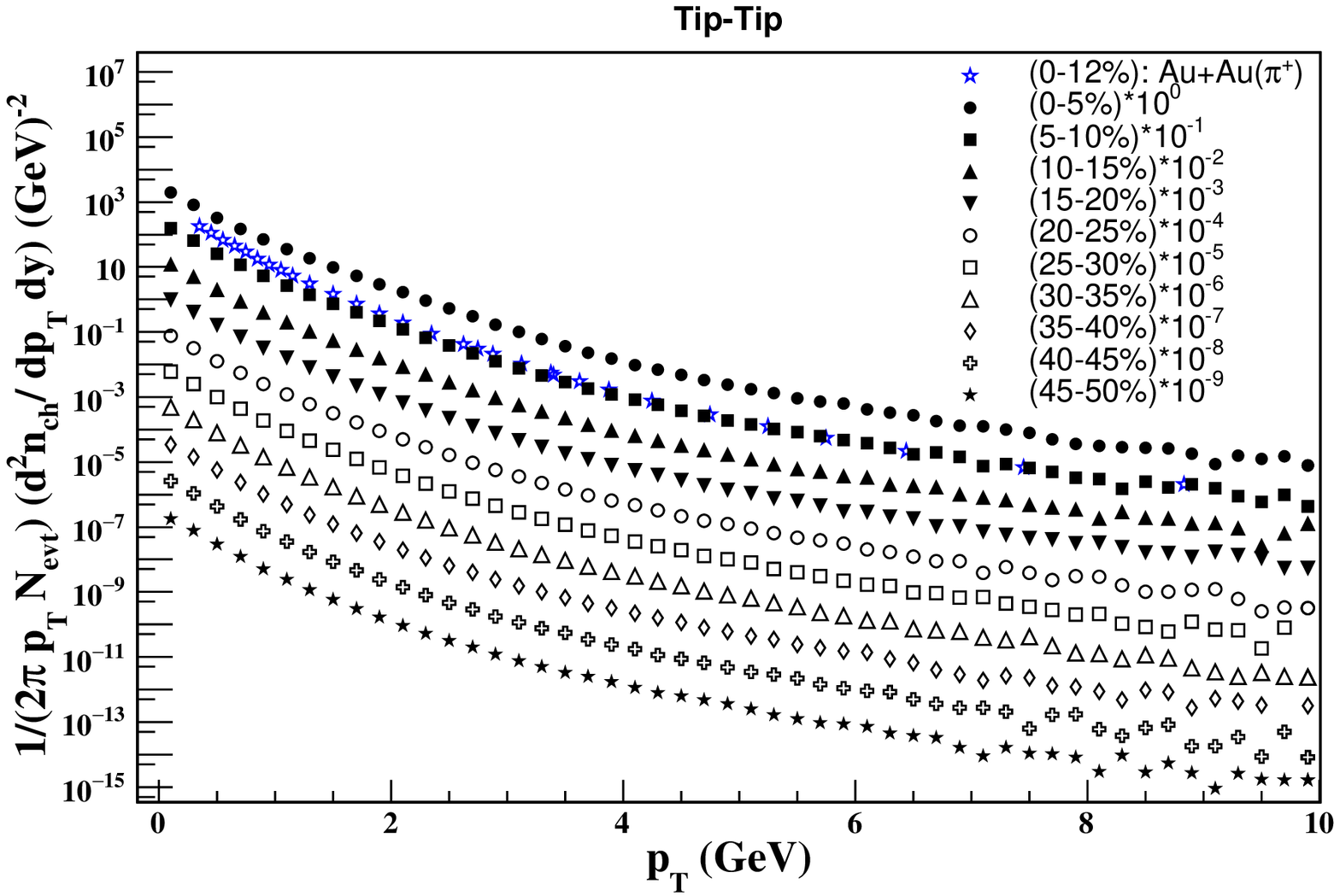}
\caption{(Color Online) Normalized transverse momentum distribution of charged hadrons in tip-tip configuration for various centrality classes.}
\end{figure}

In Fig. 15, we have demonstrated the variation of normalized transverse momentum distribution of positively charged pions ($\pi^{+}$) in central $Au+Au$ and charged pions ($\pi^{\pm}$) in  central $Pb+Pb$ collisions at $200$ GeV and $2.76$ TeV and compare HYDJET++ results with the experimental data. We have observed a suitable match between data and the model results.
Fig. 16 represents the normalized transverse momentum distribution of charged hadrons produced in $U+U$ collision in tip-tip configuration with respect to $p_{T}$ in various centrality intervals ranging from most peripheral to most central. To show the results clearly we have scaled the normalized $p_{T}$ distribution of each centrality class with different weight factors. The slope of $p_{T}$-distribution (which actually measures the inverse of source temperature from which these particles are created) increases as we move from central to peripheral collisions. This indicates that the temperature of the fireball created in central collision is higher than the peripheral collisions. We have compared the results with the experimental data of positively charged pions in central $Au+Au$ collisions at $200$ GeV of center-of-mass energy ~\cite{Abelev:2006}. One can see that the normalized transverse momentum distribution of positively charged hadrons in most central tip-tip configuration of $U+U$ collision is higher than the normalized $p_{T}$-distribution of charged pions in $Au+Au$ collisions. The central $Au+Au$ data for $0-12\%$ centrality class almost matches with the scaled $U+U$ result of $5-10\%$ centrality class at intermediate and high $p_{T}$ range.

\begin{figure}
\includegraphics[scale=0.50]{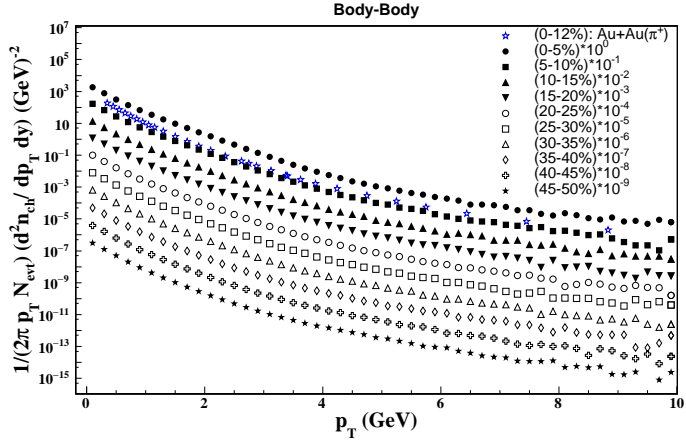}
\caption{(Color Online) Normalized transverse momentum distribution of charged hadrons in body-body configuration for various centrality classes.}
\end{figure}

\begin{figure}
\includegraphics[scale=0.50]{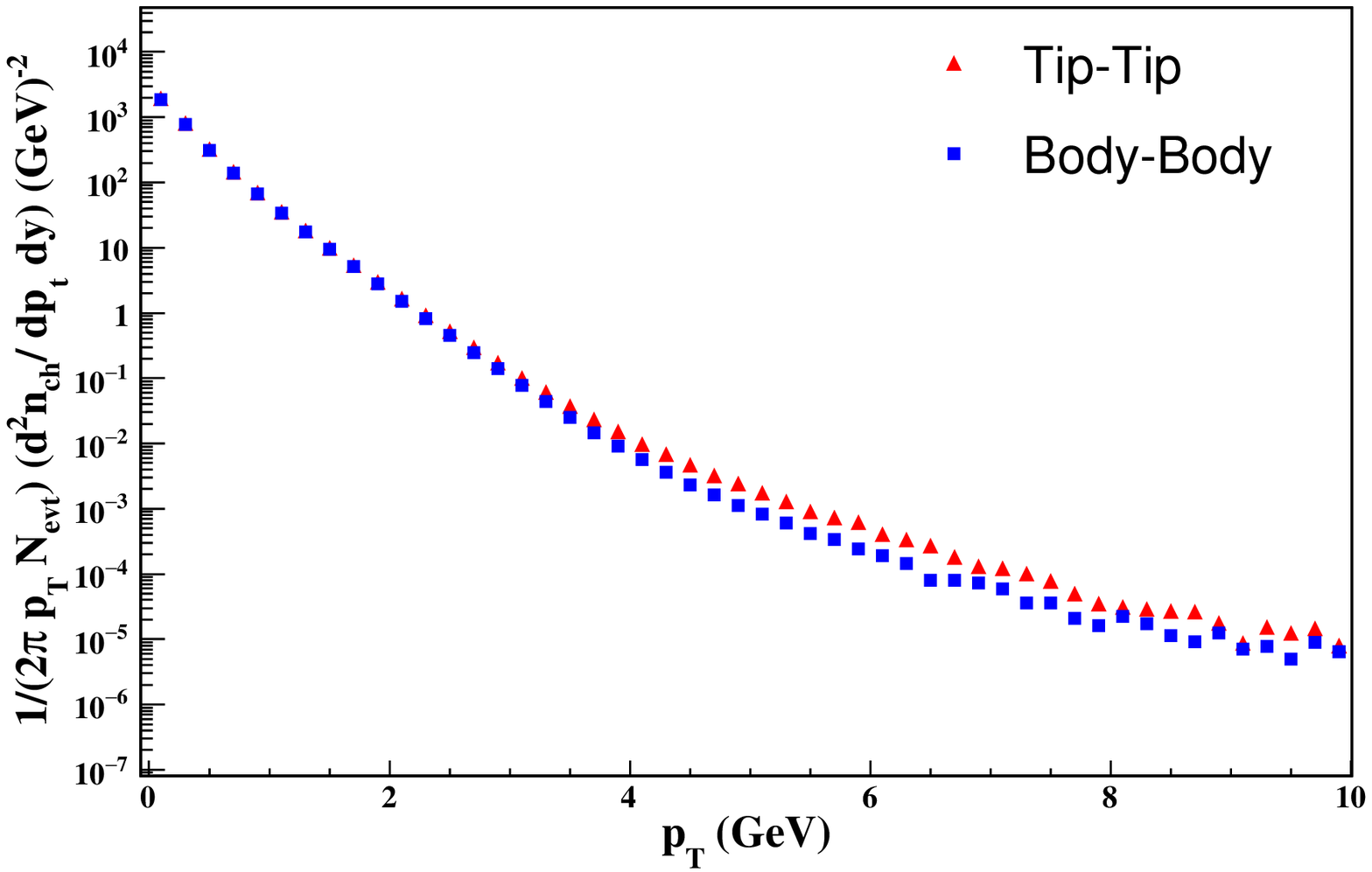}
\caption{(Color Online) Comparison of transverse momentum distribution of charged hadrons in tip-tip and body-body configuration for most central collision.}
\end{figure}

\begin{figure}
\includegraphics[scale=0.50]{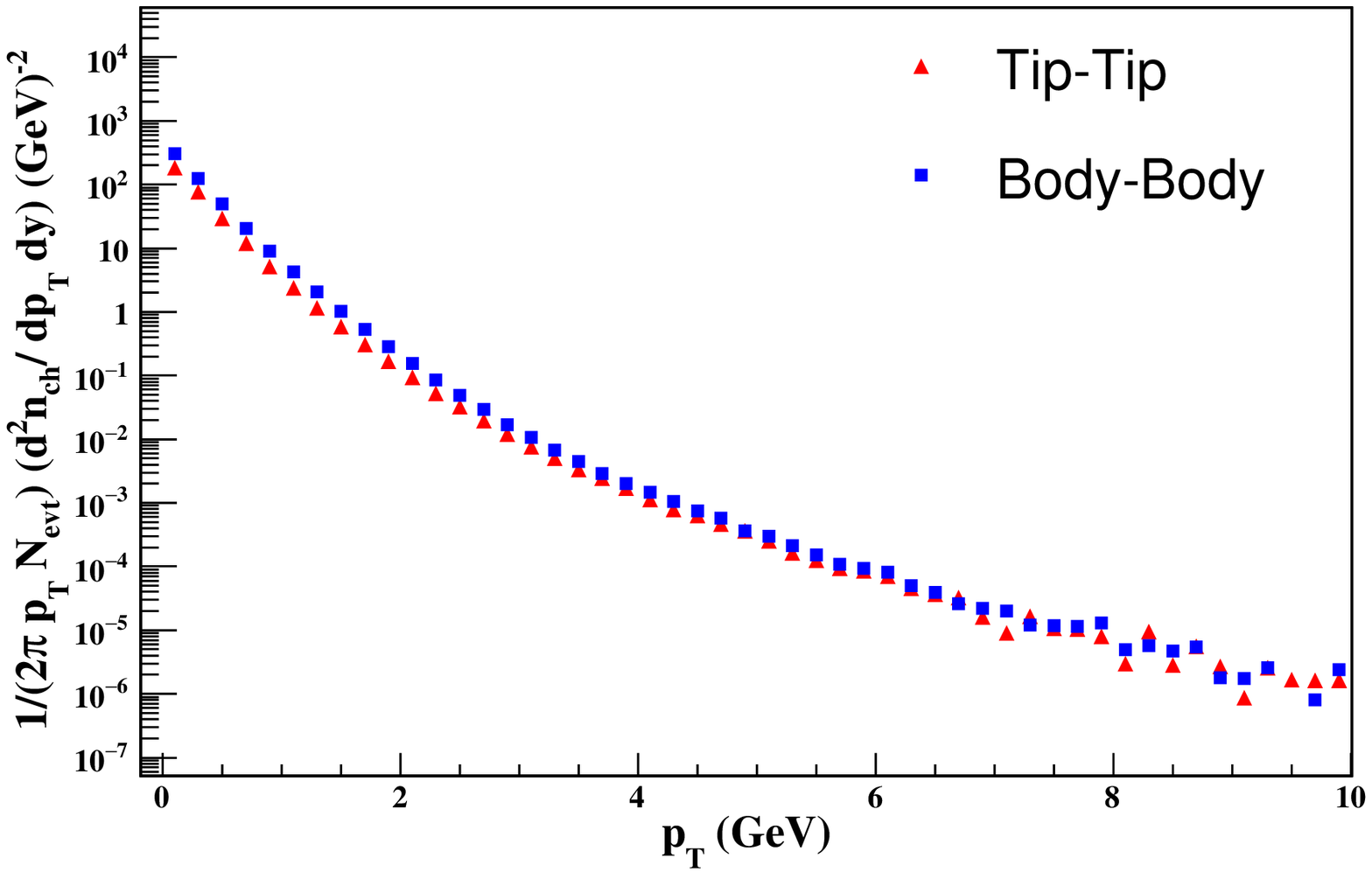}
\caption{(Color Online) Comparison of transverse momentum distribution of charged hadrons in tip-tip and body-body configuration for most peripheral collision.}
\end{figure}

In Fig. 17 we have shown the normalized transverse momentum distribution of charged hadrons in body-body configuration with respect to $p_{T}$ for various centrality classes. The central $Au+Au$ data for charged pions almost matches with the scaled $U+U$ result of $5-10\%$ centrality class at intermediate $p_{T}$ range.  However, at low and high $p_{T}$ range, the multiplicity is larger in $Au+Au$ collision than the body-body configuration of $U+U$ collision. In body-body configuration (see Fig. 18), the slope of distribution is more in comparison to tip-tip configuration for given centrality class due to the effect of transverse flow. The difference of $p_{T}$ distribution for both the configurations can be seen at intermediate and large $p_{T}$ region for central collision. As shown in Fig. 19, for most-peripheral collisions there is small difference between tip-tip and body-body configurations in low $p_{T}$ region only. As we know that most of the low-$p_{T}$ particles are due to thermal production and high-$p_{T}$ particles are due to jet fragmentation. Thus, in peripheral collision the initial configuration of nuclei affects the thermal part mostly (as shown in Fig. 12) and very small difference in jet-part but in central collisions, initial configuration mostly affect the jet-fragmentation part at higher $p_{T}$ (as shown in Fig. 11).

\begin{figure}
\includegraphics[scale=0.70]{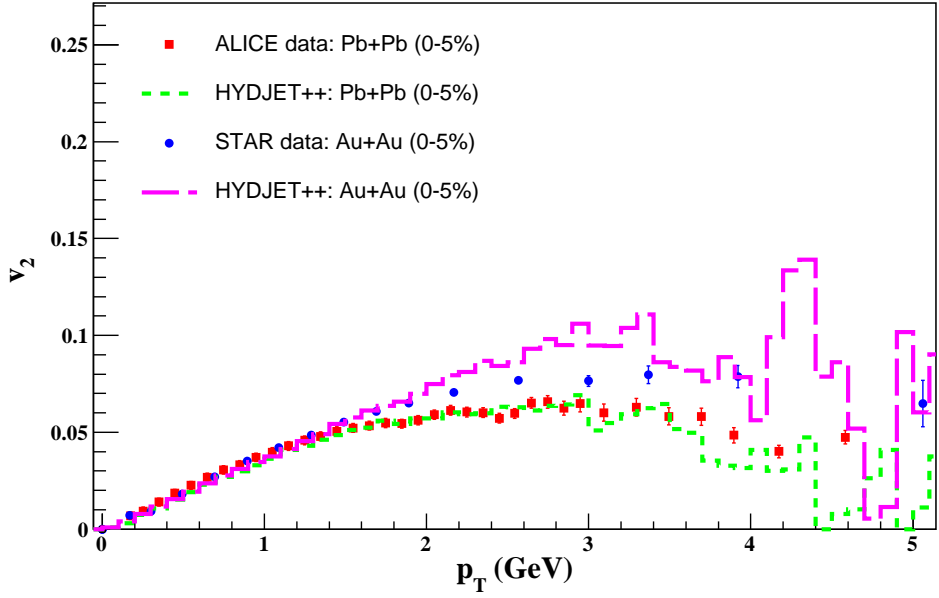}
\caption{(Color online) Variation of elliptic flow ($v_{2}$) of charged hadrons with respect to $p_{T}$ in $Au+Au$ and of charged pions in $Pb+Pb$ collisions for most central class. We have taken the experimental data from Refs.~\cite{Adams:2005,Abelev:2015}.}
\end{figure} 
\begin{figure}
\includegraphics[scale=0.50]{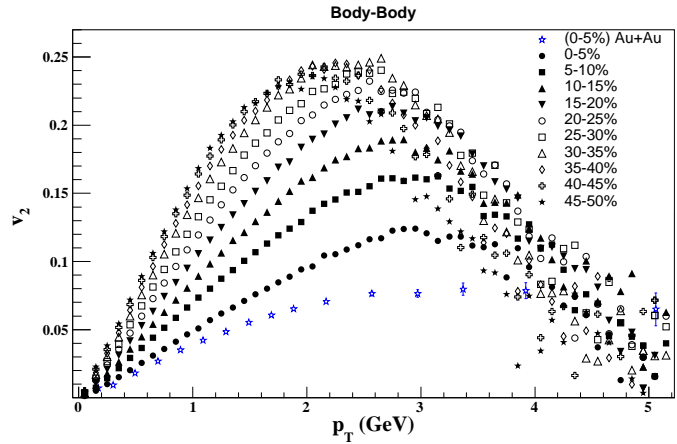}
\caption{(Color online) Variation of elliptic flow ($v_{2}$) with respect to $p_{T}$ in body-body configuration for various centrality classes.}
\end{figure}

\begin{figure}
\includegraphics[scale=0.50]{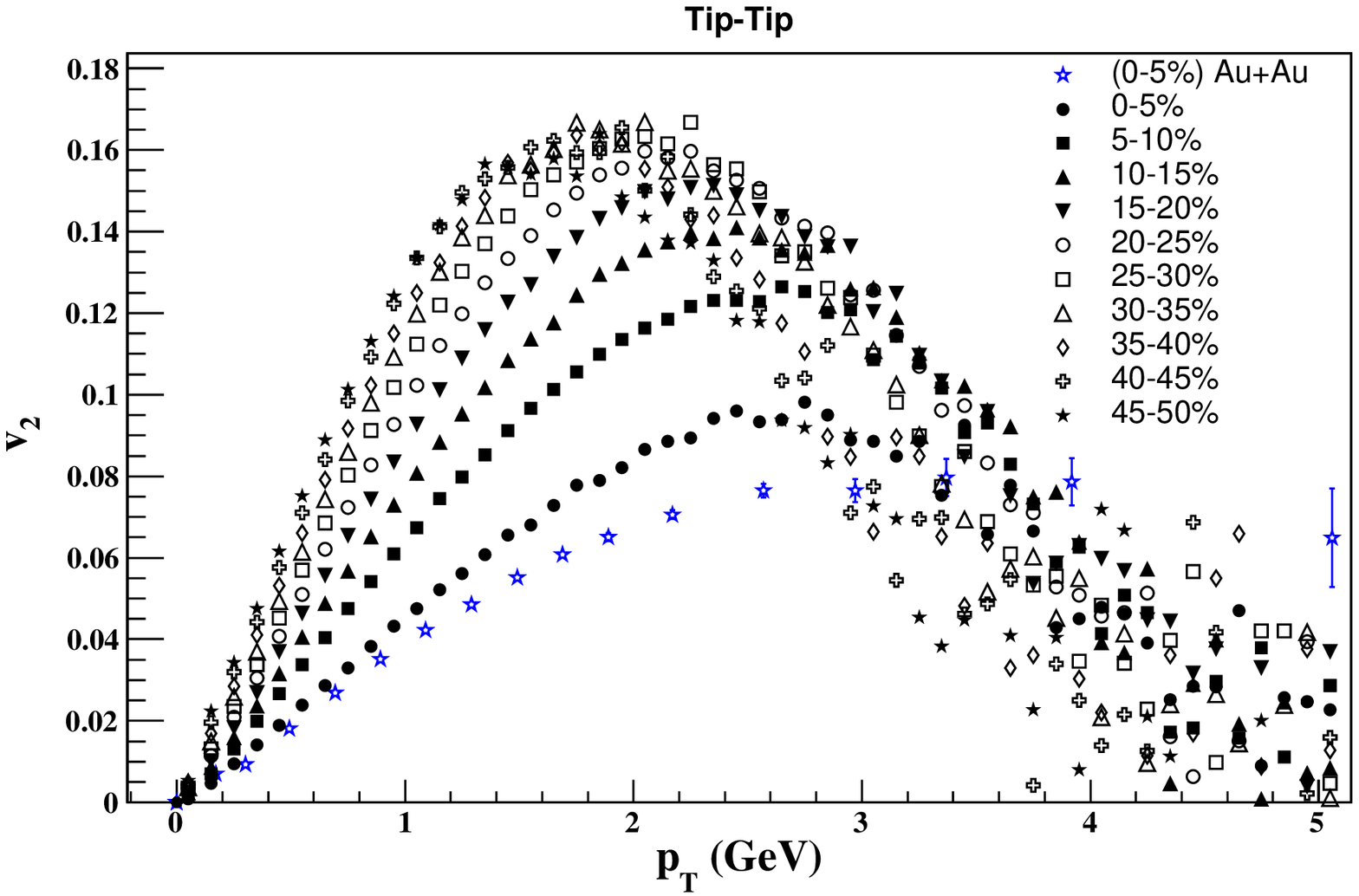}
\caption{(Color online) Variation of elliptic flow ($v_{2}$) with respect to $p_{T}$ in tip-tip configuration for various centrality classes.}
\end{figure}

In Fig. 20, we have demonstrated the variation of elliptic flow distribution of charged hadrons with respect to transverse momentum in central $Au+Au$ and of charged pions in central $Pb+Pb$ collisions at $200$ GeV and $2.76$ TeV and compare HYDJET++ results with the experimental data. We have observed a suitable match between data and the model results.
Fig. 21 demonstrates the variation of elliptic flow with respect to transverse momentum ($p_{T}$) for various centrality class in body-body configuration of $U+U$ collisions. We have shown these results for charged hadrons with $|\eta|< 0.5$. From this plot, one can observe that the elliptic flow increases with $p_{T}$ upto $p_{T}\approx 3$ GeV and then starts to decrease with further increase in $p_{T}$ for each centrality class. Further, it is clearly shown that for any given $p_{T}$ upto $3$ GeV, the elliptic flow increases as the collision becomes more and more peripheral. It is quite obvious since the initial geometrical anisotropy is very small for central collisions which actually reflects in low $v_{2}$ value for central collision. At higher $p_{T}$ the elliptic flow in each centrality class overlaps on each other. Further we have shown the elliptic flow of charged hadrons in central $Au+Au$ collision ~\cite{Adams:2005} for comparison.  We observed that $v_{2}$ in most central $Au+Au$ collision is less than $v_{2}$ in most central body-body $U+U$ collisions over the entire $p_{T}$ range considered here. Similarly, Fig. 22 presents the variation of $v_{2}$ with $p_{T}$ in different centrality class for tip-tip configurations. The qualitative behaviour of elliptic flow is quite similar to the body-body configuration. However when we see the comparison of $v_{2}$ in central $Au+Au$ data with the $v_{2}$ in tip-tip $U+U$ most central collision then one can see that $v_{2}$ of charged hadrons in $Au+Au$ collision is less than $v_{2}$ of charged hadrons in tip-tip configuration of $U+U$ most central collision.

We have shown a comparison of $v_{2}$ for tip-tip and body-body in central collisions in Fig. 23. We find that the elliptic flow of body-body configuration is slightly larger than the elliptic flow in tip-tip configuration and as we move towards larger $p_{T}$, this difference in $v_{2}$ between two configurations increases with the increase in $p_{T}$. We have also plotted the STAR experimental data of $U+U$ collisions in $0-5\%$ and $0-0.5\%$ centrality class with $|\eta|\le 1$~\cite{Pandit:2014}. 
The thought behind calculating elliptic flow for $0-0.5\%$ centrality class in STAR is that they should consists mainly tip-tip events of $U+U$ collisions. STAR collaboration has done the calculation of $v_{2}$ for charged hadrons. By comparison we observe that STAR data of $0-0.5\%$ centrality class have lower $v_{2}$ in comparison to our tip-tip as well as body-body results. However $0-0.5\%$ centrality class data matches with our tip-tip results when $p_{T}<1$ GeV (see inset of Fig. 23). Further the experimental data from $0-5\%$ is greater than our tip-tip result when $p_{T}<1$ GeV but it matches with our tip-tip results for intermediate and large $p_{T}$. Another important observation is that the $v_{2}$ in body-body configuration is higher than both these data sets along with tip-tip results from HYDJET++. In peripheral collisions (see Fig. 24), the qualitative difference between $v_{2}$ in tip-tip and body-body configuration is same. However the magnitude of difference in $v_{2}$ of charged hadrons is quite visible even at $p_{T}\approx 1$ and as we move towards intermediate $p_{T}$ this difference increases in same manner as in the case of Fig. 23. Further, in intermediate region the magnitude of difference is quite large in the case of most peripheral collision than the most central $U+U$ collisions. Another observation is that the maxima in $v_{2}$ is at $p_{T}\approx 3$ in central collision while it comes down to $p_{T}\approx 2$ in peripheral collision.

\begin{figure}
\includegraphics[scale=0.50]{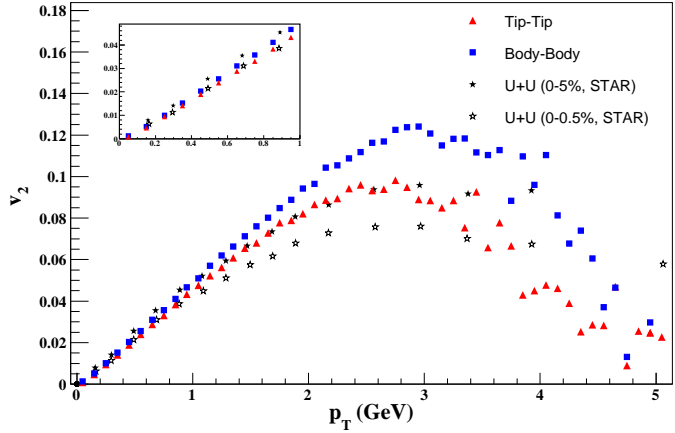}
\caption{(Color Online) Comparison of $v_{2}$ of charged hadrons with respect to $p_{T}$ in body-body and tip-tip configuration for most central collision. STAR data is taken from Ref.~\cite{Pandit:2014}.}
\end{figure}

\begin{figure}
\includegraphics[scale=0.50]{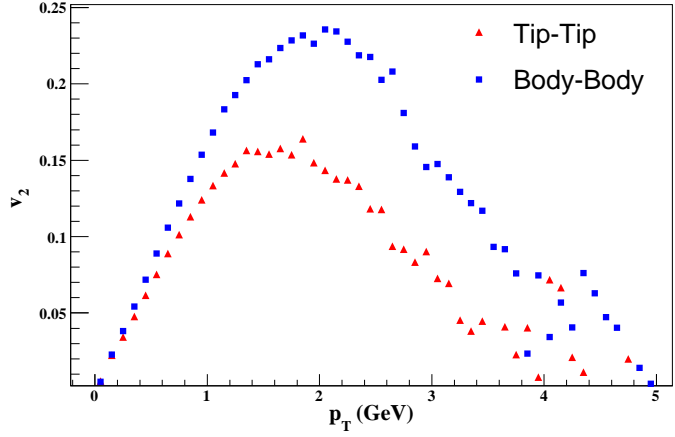}
\caption{(Color Online) Comparison of $v_{2}$ of charged hadrons with respect to $p_{T}$ in body-body and tip-tip configuration for most peripheral collision.}
\end{figure}

\begin{figure}
\includegraphics[scale=0.50]{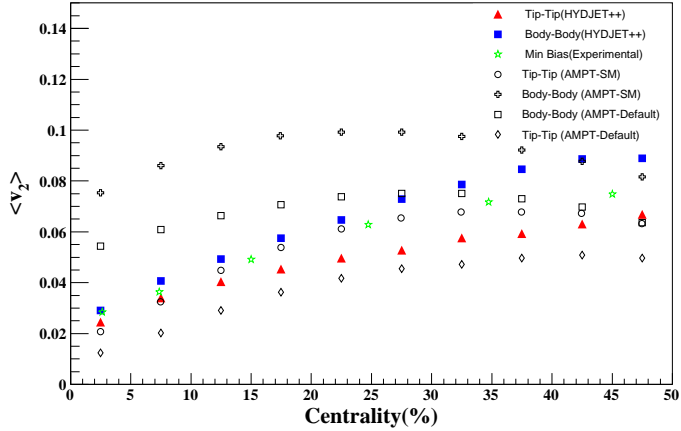}
\caption{(Color online) Comparison of $v_{2}$ of charged hadrons with respect to centrality in body-body and tip-tip configuration. Green stars are experimental data taken from Ref. ~\cite{Pandit:2014} and AMPT model data is taken from Ref. ~\cite{Rihan:2012}.}
\end{figure}

In Fig. 25, we have shown the effect of centrality on mean elliptic flow for body-body and compared them with the corresponding results of tip-tip configurations. We have integrated over $p_{T}$ from $0.001$ GeV to $5$ GeV. From this result one can see that $\langle v_{2}\rangle$ increases in going from central to peripheral which is actually due to a increase in eccentricity going from central to peripheral collisions. However from here it is clear that in central collisions the difference in magnitude between body-body and tip-tip collisions is small. However, in semi-peripheral as well as in peripheral collisions, one can distinguish between body-body and tip-tip events by observing the $\langle v_{2}\rangle$ magnitude of charged hadrons. We have also shown the results obtained in Ref. ~\cite{Rihan:2012} using AMPT model in two different modes (string melting mode and default mode). We found that the qualitative behaviour of variation of $\langle v_{2}\rangle$ with centrality in HYDJET++ is quite opposite to AMPT model and shows a small difference in $v_{2}$ for tip-tip and body-body configuration in central events and a large difference in peripheral events. On the other side AMPT has shown opposite behaviour. We have also plotted the STAR experimental data ~\cite{Pandit:2014} of $v_{2}$ as a function of centrality for minimum bias events. We found that the experiment results are nearly in between the HYDJET++ model results for tip-tip and body-body configurations. However, the experimental data is between the AMPT-Default mode results for tip-tip and body-body configurations in central and mid-central events but not in peripheral events. In AMPT-SM mode, the experimental data is very close to tip-tip configuration results.

In summary, we have calculated and shown the pseudorapidity density and transverse momentum distributions of charged hadrons produced in $U+U$ collisions at $\sqrt{s_{NN}}=193$ GeV in various initial geometrical configurations. In present study, it has been shown that the correlation between multiplicity and initial geometrical configurations of $U+U$ collisions is small which is in accordance with the recent experimental observation. However, the experimental results are quite preliminary due to complexity in disentangling the tip-tip and body-body events. We have shown the midrapidity charged-particle multiplicity distribution from HYDJET++ model which is in good agreement with the experimental results for minimum bias events. Further, we have shown the evolution of elliptic flow with $p_{T}$ and centrality in different configurations of $U+U$ collisions. It has been observed that elliptic flow generated in body-body collisions is larger than tip-tip collisions but the difference in magnitude of $v_{2}$ is small in central collisions and large in peripheral events. Further, we found that our tip-tip results of elliptic flow matches with STAR experiment result of 0-0.5\% centrality class when $p_{T}<1$. At last, we have observed that the experimental results of $v_{2}$ as a function of centrality for minimum bias events are nearly in between the tip-tip and body-body configuration results of our model. However, this is not the case for AMPT results. Finally we may conclude that our present study will shed some light on the particle production mechanism and the evolution of the fireball created in various geometrical configurations of $U+U$ collisions specially the entanglement of hard (jet) and soft (hydro) part in body-body and tip-tip configurations.

\section{Acknowledgments}
\noindent 
Authors gratefully acknowledge personal communications and the useful comments/suggestions on the present manuscript by Prof. I.P. Lokhtin and his group. PKS is grateful to IIT Ropar, India for providing an institute postdoctoral research grant. PKS is also thankful for the pleasant stay at Department of Physics, BHU where work was done. OSKC would like to thank Council of Scientific and Industrial Research (CSIR), New Delhi for providing a research fellowship. AS acknowledges the financial support obtained from UGC under research fellowship scheme in central universities. This research work is supported partially by DST FIST, DST PURSE and UGC-CAS programmes.

\pagebreak


\begin{thebibliography}{40}
\bibitem{Schenke:2012}
  B. Schenke, P. Tribedy, and R. Venugopalan,
  Phys.\ Rev.\ Lett.\ {\bf 108}, 252301 (2012);

\bibitem{Eskola:2000}
  K. J. Eskola, K. Kajantie, P. V. Ruuskanen, and K. Tuominen,
  Nucl. Phys. {\bf B 570}, 379 (2000).
\bibitem{Eskola:2002}
  K. J. Eskola,
  Nucl. Phys. {\bf A 698}, 78 (2002).

\bibitem{Kharzeev:2004}
  D. Kharzeev, E. Levin, and M. Nardi,
  Nucl. Phys. {\bf A 730}, 448 (2004);
  Nucl. Phys. {\bf A 747}, 609 (2005).

\bibitem{Kharzeev:2001}
  D. Kharzeev, and M. Nardi,
  Phys. Lett. {\bf B 507}, 121 (2001);
  D. Kharzeev, E. Levin,
  Phys. Lett. {\bf B 523}, 79 (2001).

\bibitem{Schee:2013}
  W. van der Schee, P. Romatschke, and S. Pratt,
  Phys.\ Rev.\ Lett.\ {\bf 111}, 222302 (2013).

\bibitem{Berges:2014}
  J. Berges, B. Schenke, S. Schlichting, and R. Venugopalan,
  Nucl. Phys. {\bf A 931}, 348 (2014).

\bibitem{Kurkela:2014}
  A. Kurkela, and E. Lu,
  Phys.\ Rev.\ Lett.\ {\bf 113}, 182301 (2014), arXiv:1405.6318 [hep-ph].

\bibitem{Miller:2007}
  M. L. Miller, K. Reygers, S. J. Sanders, and P. Steinberg,
  Ann.\ Rev.\ Nucl.\ Part.\ Sci.\ {\bf 57}, 205 (2007), arxiv:nucl-ex/0701025.

\bibitem{Lin:2005}
Z. W. Lin, C. M. Ko, B. A. Li, B. Zhang, and S. Pal,
Phys. Rev. {\bf C 72}, 064901 (2005).

\bibitem{Bleicher:1999}
M. Bleicher et al.,
J. Phys. {\bf G 25}, 1859 (1999).

\bibitem{Pandit:2014}
  Yadav Pandit (for the STAR collaboration),
  arXiv:1405.5510 [nucl-ex] (2014).

  \bibitem{Bjorn:1403} 
B. Schenke, P. Tribedy, and R. Venugopalan, 
Phys. Rev {\bf C 89}, 064908 (2014).

\bibitem{Sergei:1006}
S. A. Voloshin,
Phys. Rev. Lett. {\bf 105}, 172301 (2010).

\bibitem{Hiroshi:2009} 
H. Masui, B. Mohanty, and Nu Xu, 
Phys.\ Lett.\ {\bf B 679}, 440–444 (2009).

\bibitem{Shou:2015} 
Q. Y. Shou, Y. G. Ma, P. Sorensen, A. H. Tang, F. Videb$\ae$k, and H. Wang, 
Phys.\ Lett.\ {\bf B 749}, 215-220 (2015).

\bibitem{Gold:2015}
A. Goldschmidt, Z. Qiu, C. Shen, and U. Heinz, 
Phys.\ Rev.\ {\bf C 92}, 044903 (2015).

\bibitem{Hirano:2011}
T. Hirano, P. Huovinen, and Y. Nara,
Phys. Rev. {\bf C 83}, 021902(R) (2011).

\bibitem{Kuhlman:2005}
A. Kuhlman, and U. Heinz,
Phys. Rev. {\bf C 72}, 037901 (2005).

\bibitem{John:1311} 
J. Bloczynski, Xu-G. Huang, X. Zhang, and J. Liao, 
Nucl. Phys. {\bf A 939}, 85 (2015).

\bibitem{Chatterjee:2016}
S. Chatterjee, S. K. Singh, S. Ghosh, Md. Hasanujjaman, J. Alam, and S. Sarkar,
Phys. Lett. {\bf B 758}, 269 (2016).

\bibitem{Rybcz:2013}
M. Rybczynski, W. Broniowski, and G. Stefanek,
Phys. Rev. {\bf C 87}, 044908 (2013).

\bibitem{Moreland:2015}
J. S. Moreland, J. E. Bernhard, and S. A. Bass,
Phys. Rev. {\bf C 92}, 011901 (2015).

\bibitem{Eremin:2003}
S. Eremin, and S. Voloshin,
Phys. Rev. {\bf C 67}, 064905 (2003).

\bibitem{Adler:2014}
S. S. Adler et al., PHENIX Collaboration, 
Phys. Rev. {\bf C 89}, 044905 (2014).

\bibitem{Lokhtin:2009}
  I. P. Lokhtin, L. V. Malinina, S. V. Petrushanko, A. M. Snigirev, I. Arsene, and K. Tywoniuk,
  Comput.\ Phys.\ Commun.\ {\bf 180}, 779-799 (2009).

\bibitem{Bravina:2017}
  L. V. Bravina, I. P. Lokhtin, L. V. Malinina, S. V. Petrushanko, A. M. Snigirev, and E.E. Zabrodin,
  Eur.\ Phys.\ J.\ {\bf A 53}, 219 (2017).

\bibitem{Lokhtin:2006}
  I. P. Lokhtin, and A. M. Snigirev,
  Eur.\ Phys.\ J.\ {\bf C 45}, 211 (2006).

\bibitem{Lokhtin:2708}
  I. P. Lokhtin, S. V. Petrushanko, A. M. Snigirev, and K. Tywoniuk,
  PoS (LHC07) 003 {arXiv:0809.2708}.

\bibitem{Lokhtin:2082}
  I. P. Lokhtin, L. V. Malinina, S. V. Petrushanko, A. M. Snigirev, I. Arsene, and K. Tywoniuk,
  PoS (LHC08) 002 [arXiv:0810.2082].

\bibitem{Anderson:1998}
  B. Andersson, The Lund Model, Cambridge University Press (1998).

\bibitem{Gribov:1969}
  V. N. Gribov,
  Sov.\ Phys.\ JETP {\bf 29}, 483 (1969).

\bibitem{sve}
B. Svetitsky, 
Phys. Rev. {\bf D 37}, 2484 (1988).

\bibitem{pks_drag}
P. K. Srivastava, and B. K. Patra,
Eur. Phys. J. {\bf A 53}, 116 (2017).

\bibitem{Bjorken:1983}
  J. D. Bjorken,
  Phys.\ Rev.\ {\bf D 27}, 140 (1983).

  \bibitem {Rihan:2012}
  Md. Rihan Haque, Zi-Wei Lin, and Bedangadas Mohanty,
  Phys.\ Rev.\ {\bf C 85}, 034905 (2012).


\bibitem{OSK:2016}
  O. S. K. Chaturvedi, P. K. Srivastava, Ashwini Kumar, and B. K. Singh,
  Eur.\ Phys.\ J.\ Plus {\bf  131}, 438 (2016).

\bibitem{Loizides:2549}
  C. Loizides, J. Nagle, and P. Steinberg,
  SoftwareX {\bf 1-2}, 13 (2015).

\bibitem{Lokhtin:2000}
  I. P. Lokhtin, and A. M. Snigirev,
  Eur.\ Phy.\ J. \ {\bf C 16}, 527 (2000).



  
\bibitem{Amelin:2006}
  N. S. Amelin {\it et al.},
  Phys.\ Rev.\ {\bf C 74}, 064901 (2006).

\bibitem{Amelin:2008}
  N. S. Amelin {\it et al.},
  Phys.\ Rev.\ {\bf C 77}, 064901 (2008).

\bibitem{cleymans}
J. Cleymans, H. Oeschler, K. Redlich, and S. Wheaton, Phys.
Rev. {\bf C 73}, 034905 (2006); J. Cleymans et al., Phys. Lett. {\bf B 660},
172 (2008).

\bibitem{skt}
S. K. Tiwari, P. K. Srivastava, and C. P. Singh, Phys. Rev. {\bf C 85}, 014908 (2012).

\bibitem{sandeep}
S. Chatterjee, S. Das, L. Kumar, D. Mishra, B. Mohanty, R. Sahoo , and N. Sharma, AHEP {\bf 2015}, 349013 (2015).

\bibitem{andronic}
A. Andronic, P. Braun--Munzinger, and J. Stachel, arXiv:nucl-th/0511071v3 (2006).

\bibitem{Lokesh}
L. Kumar (for STAR Collaboration), Cent. Eur. J. Phys. {\bf 10(6)}, 1274 (2012).

\bibitem{Torrieri:2005}
  G. Torrieri {\it et al.},
  Comput.\ Phys.\ Commun.\ {\bf 167}, 229 (2005).



\bibitem{Eyyubova:2009}
  G. Eyyubova {\it et al.},
  Phys. Rev. {\bf C 80}, 064907 (2009).

\bibitem{Retiere:2004}
  F. Retiere, and M. A. Lisa,
  Phys. Rev. {\bf C 70}, 044907 (2004).

\bibitem{Huovinen:2001}
  P. Huovinen, P. F. Kolb, U. Heinz, P. V. Ruuskanen, and S. A. Voloshin,
  Phys. Lett. {\bf B 503}, 58 (2001).

\bibitem{Wiedemann:1998}
  U. A. Wiedemann,
  Phys. Rev. {\bf C 57}, 266 (1998).

\bibitem{Broniowski:2003}
  W. Broniowski, A. Baran, and W. Florkowski,
  AIP Conf. Proc. {\bf 660}, 185 (2003).

\bibitem{Adare:2016}
  A. Adare {\it et al.} [PHENIX Collaboration],
  Phys.\ Rev.\ {\bf C 93}, 024901 (2016).
\bibitem{Abbas:2013}
  E. Abbas {\it et al.} [ALICE Collaboration],
  Phys.\ Lett.\ {\bf B 726}, 610 (2013).
  
\bibitem{Back:2003}
  
  B. B. Back {\it et al.} [PHOBOS Collaboration],
  Phys.\ Rev.\ Lett.\ {\bf 91}, 052303 (2003).


  
\bibitem{Abelev:2006}
  B. I. Abelev {\it et al.} [STAR Collaboration],
  Phys.\ Rev.\ Lett.\ {\bf 97}, 152301 (2006).
\bibitem{Abelev:2014}
  B. Abelev {\it et al.} [ALICE Collaboration],
  Phys.\ Lett.\ {\bf B 736}, 196 (2014).
  
\bibitem{Adams:2005}
  J. Adams {\it et al.} [STAR Collaboration],
  Phys.\ Rev.\ {\bf C 72}, 014904 (2005).
  
\bibitem{Abelev:2015}
  B. Abelev {\it et al.} [ALICE Collaboration],
  J.\ High\ Eenergy\ Phys.\ {\bf 2015}, 190 (2015).  

\end{thebibliography}
\end{document}